\documentclass[aps,pre,superscriptaddress,showpacs,amsfonts]{revtex4}
\usepackage{graphicx}
\usepackage{bm}

\newcommand{\bgamma}{{\bm{\gamma}}}
\newcommand{\bsigma}{{\bm{\sigma}}}

\begin{document}

\title{Partially fluidized shear granular flows: Continuum theory and MD
simulations}

\author{Dmitri Volfson}
\affiliation{Institute for Nonlinear Science, University of California,
San Diego, La Jolla, California 92093-0402 }
\author{Lev S. Tsimring}
\affiliation{Institute for Nonlinear Science, University of California,
San Diego, La Jolla, California 92093-0402 }
\author{Igor S. Aranson}
\affiliation{Argonne National Laboratory,
9700 South  Cass Avenue, Argonne, Illinois 60439}

\date{\today}

\begin{abstract} 
The continuum theory of partially fluidized 
shear granular flows is tested and calibrated using 
two dimensional  soft particle molecular   dynamics simulations. 
The theory is based on the relaxational dynamics of the order parameter
that describes the transition between static and flowing regimes of
granular material.  We define the order
parameter as a fraction of static contacts among
all contacts between particles. We also propose and verify by direct
simulations the constitutive relation based on the splitting of the
shear stress tensor into a``fluid part''
proportional to the strain rate tensor, and a remaining ``solid part''.
The ratio of these two parts is a function of the order parameter. 
The rheology of the fluid component agrees well with the kinetic theory of
granular fluids even in the dense regime.  
Based on the hysteretic bifurcation diagram for a
thin shear granular layer obtained in simulations, we construct the ``free
energy'' for the order parameter. The theory calibrated using
numerical experiments with the thin granular layer is applied to 
the surface-driven stationary two dimensional  granular flows in a thick granular layer
under gravity. 
\end{abstract} 

\date{\today}

\maketitle

\section{Introduction}

In the last few year there have been many experimental
\cite{behringer99,gollub98,gollub00,chicago,pouliquen99,daerr,Tsai02} and theoretical
\cite{grest91,wolf96,grest02,sparks99,sparks02,lemaitre,ertas} studies that
explored a broad range of granular flow
conditions from rapid dilute flows to slow dense flows, as well as the details
of the shear-driven fluidization transition. 
While dilute granular flows can be well described by the
kinetic theory of dissipative granular gases \cite{Jenkins85}, dense granular flows
still present significant difficulty in formulation of a 
continuous theory. 
In Ref. \cite{savage98}, Savage proposed a continuum theory for slow
dense granular flows based on the 
so-called associated flow rule that relates the strain rate and the shear stress 
in plastic frictional systems.  Averaging strain rate fluctuations
yields a Bingham-like constitutive relation in which the shear stress has
a viscous as well as a strain-rate independent parts. 
According to this
theory, the stress and strain rate tensors are always co-axial and,
furthermore, it also postulates that the viscosity diverges as the
density approaches close packing limit. 
Losert et al. \cite{gollub00} (see also \cite{bocquet02}) proposed a
similar hydrodynamic model based on a Newtonian stress-strain
constitutive relation with density dependent viscosity without
strain-rate independent component.  As observed in Ref. \cite{gollub00},
the ratio of the full shear stress to the strain rate diverges at the
fluidization threshold. This was also interpreted in
Ref. \cite{gollub00} as a divergence of the viscosity coefficient when
the volume fraction approaches the randomly packed limit.  
This description works only in a fluidized state and can not properly 
account for hysteretic phenomena in which 
static and fluidized states co-exist under the same
external load, such as stick-slip oscillations
\cite{gollub98}, avalanching \cite{daerr}, or shear band formation. 

In many granular flows of interest static and dynamics regions co-exist
under the same external load conditions.
Examples of such hysteretic phenomena include stick-slip oscillations
\cite{gollub98}, avalanching \cite{daerr}, or shear band formation. 
This calls for a unified theory which would be applicable
both in the flowing regime and in the static regime.  

In our recent papers \cite{AT1,AT2} we proposed a different approach based on
the order parameter description of the granular matter. The value of the
order parameter specifies the ratio between static and
fluid parts of the stress tensor. The order parameter was assumed
to obey dissipative dynamics governed by a  free energy
functional with two local minima.  This description based on the separation of static 
and fluid components of the shear stress, calls for an alternative 
definition of viscosity as a ratio
of the {\em fluid} part of the shear stress to the strain rate. Since
the fluid shear stress vanishes together with the strain rate, the 
viscosity coefficient in our theory is expected to remain finite at
the fluidization threshold. We assumed the simplest Newtonian friction
law, so the viscosity coefficient is a constant.
 This model yielded a good qualitative description of many
phenomena occurring in granular flows, such as hysteretic transition to
chute flow, stick-slip regime of a driven near-surface flow, structure
of avalanches in shallow chute flows, etc. 

However, several important issues have not been addressed: mesoscopic
definition of the order parameter, quantitative specification of the
order parameter dynamics and the constitutive relation.  In this paper
we set out to perform two dimensional (2D) molecular dynamics simulations which should
provide us with the way to achieve these goals. We classify all contacts
as either ``fluid-like" or ``solid-like" and define the order parameter
as a mesoscopic space-time average fraction of solid-like contacts.
Using this order parameter we obtain the constitutive relation and the relaxational
dynamics of the order parameter directly from simulations of granular
flow in a thin Couette geometry at zero gravity. 
Preliminary account of our results is presented in Ref. \cite{volfson}. 

The paper is organized
as follows. Section \ref{sechydro} outlines the standard granular
hydrodynamics theory based on the kinetic theory of dissipative granular
gases. Section \ref{sec:order parameter} introduces the continuum description of
partially fluidized flows based on the relaxational dynamics of the order parameter.
In Section \ref{sec:mds} we describe our 2D molecular dynamics simulations
and define measurement protocol for the order parameter and fluid and
solid components of the stress tensor.  In Section \ref{sec:testbed} we
study Couette flow in a thin granular layer to obtain the free energy
controlling the relaxational order parameter dynamics and  to extract the
constitutive relations.  In Section \ref{sec:deep} the obtained set of
equations is used to calculate the stress and velocity distributions in
a {\em different} system, a thick granular layer under non-zero gravity
driven by a moving heavy upper plate.

\section{Granular hydrodynamics
\label{sechydro}} 
In this Section we outline the standard continuum description of
granular flows based on continuity equations for mass, momentum and
fluctuation kinetic energy (or ``granular temperature''). This description 
is usually applied to dilute granular gases 
where it can be rigorously derived from the kinetic theory
\cite{Jenkins85}, although  slightly modified 
hydrodynamics based on kinetic  theory often works reasonably well  for 
relatively dense flows,  
even though the kinetic theory itself is not applicable to these conditions. 

The mass, momentum and energy conservation equations have the usual form
\begin{eqnarray}
\frac{D{\bf \nu}}{Dt}&=&-\nu \nabla\cdot {\bf u},
\label{mass}\\
\nu\frac{D{\bf u}}{Dt}&=&-\nabla\cdot \bsigma +\nu {\bf g},
\label{momentum}\\
\nu\frac{DT}{Dt}&=&-\bsigma : \dot{\bf \bgamma} -
\nabla\cdot {\bf q}- \varepsilon,
\label{energy}
\end{eqnarray}
where $\nu$ is the density, ${\bf u}$ is the velocity field, 
 $T=\langle \tilde{u}_i\tilde{u}_i\rangle/2$ is the granular temperature,
$D/Dt=\partial_t+({\bf u}\cdot\nabla)$ is the material derivative,
${\bf g}$ is the gravity acceleration, $\sigma_{\alpha \beta}$ is the stress
tensor,  ${\bf q}$ is the energy flux vector,
$\dot\gamma_{\alpha \beta}=\partial_\alpha u_\beta+\partial_\beta u_\alpha$ 
is the strain
rate tensor, and $\varepsilon$ is the energy dissipation rate. 

These three equations have to be supplemented by the
constitutive relations for the stress tensor $\bsigma$, energy flux ${\bf q}$, and
the energy dissipation rate $\varepsilon$. For dilute systems, a linear
relations between stress $ \bsigma$ and strain rate $ \dot  \bgamma$ is obtained, 
\begin{eqnarray}
\sigma_{\alpha \beta}&=&[p+(\mu-\lambda)\mbox{Tr}\dot\bgamma]\delta_{\alpha \beta}-\mu \dot\gamma_{\alpha \beta},
\label{c1}\\
{\bf q}&=&-\kappa \nabla T,
\label{c2}
\end{eqnarray}
In the kinetic theory of granular gases \cite{Jenkins85}, these equations are closed with
the following equation of state
\begin{equation}
p=\frac{4\nu T}{\pi d^2}[1+(1+e)G(\nu)],
\label{state}
\end{equation}
and the expressions for the shear and bulk viscosities 
\begin{eqnarray}
\mu&=&\frac{\nu T^{1/2}}{2\pi^{1/2} d G(\nu)}
\left[1+2G(\nu)+\left(1+\frac{8}{\pi}\right)G(\nu)^2\right],
\label{shear}\\
\lambda&=&\frac{8\nu G(\nu)T^{1/2}}{\pi^{3/2} d},
\label{bulk}
\end{eqnarray}
the thermal conductivity
\begin{equation}
\kappa=\frac{2\nu T^{1/2}}{\pi^{1/2} d G(\nu)}
\left[1+3G(\nu)+\left(\frac{9}{4}+\frac{4}{\pi}\right)G(\nu)^2\right],
\label{kappa}
\end{equation}
and the energy dissipation rate
\begin{equation}
\varepsilon=\frac{16\nu G(\nu)T^{3/2}}{\pi^{3/2} d^3}(1-e^2).
\label{eps}
\end{equation}
Here $0<e<1$ is restitution coefficient and $d$ is particle diameter.
The function $G(\nu)$ which enters these relations is the spatial
particle-particle correlation function, and for a dilute 2D gas of elastic 
hard disks was derived by Carnahan and Starling \cite{carnahan},
\begin{equation}
G_{CS}(\nu)=\frac{\nu(1-7\nu/16)}{(1-\nu)^2}
\label{gCS}
\end{equation}
This formula is expected to work for densities roughly below 0.7. For
high density granular gases, this function has been calculated 
using free volume theory \cite{kirkwood51}, 
\begin{equation}
G_{FV}=\frac{1}{(1+e)\left[(\nu_c/\nu)^{1/2}-1\right]}
\label{gFV}
\end{equation}
where $\nu_c\approx 0.82$ is the density of the random close packing limit.
Luding \cite{Luding01} proposed a global fit
\begin{equation}
G_{L}=G_{CS}+(1+\exp(-(\nu-\nu_0)/m_0))^{-1})(G_{FV}-G_{CS})
\label{gL}
\end{equation}
with empirically fitted parameters $\nu_0\approx0.7$ and
$m_0\approx10^{-2}$.  However, even with this extension, the continuum
theory comprised of Eqs.(\ref{mass})-(\ref{eps}) cannot describe the
force chains which transmit stress via persistent contacts remaining in
the dense granular flows, as well as the transition from solid to static
regimes and coexisting solid and fluid phases.

\section{Order parameter description of partially fluidized granular flows 
\label{sec:order parameter}}

In this Section we review briefly our continuum theory \cite{AT1,AT2}
which provides an alternative approach to the formulation of the constitutive 
relations in partially fluidized granular flows. In the dense flow
regime, the granular matter can be considered incompressible, so
Eq. (\ref{mass}) can be replaced by $\nabla\cdot {\bf u}=0$ and the
density $\nu=\nu_c$. This also allows us to drop the energy equation
(\ref{energy}) and the equation of state
(\ref{state}), as for $\nu\to\nu_c$, $G(\nu)\to\infty$ and $T\to0$, so
that $G(\nu)T\to const$. This of course leads to the familiar 
divergence of the viscosity coefficient $\mu\propto G(\nu)T^{1/2}$.

Next, we separate the stress tensor $\bsigma$ into two parts, a static (contact) part
$\bsigma^s$, and a fluid part $\bsigma^f$. The latter is assumed to take a
purely Newtonian form
\begin{equation}
\sigma^f_{\alpha \beta}=p_f\delta_{\alpha \beta}-\mu_f\dot\gamma_{\alpha
\beta}
\label{sigmaf}
\end{equation}
where $p_f$ is the ``partial'' fluid pressure, $\mu_f$ is the viscosity coefficient 
associated with the fluid stress tensor which is {\em different} from $\mu_0$ 
introduced for the full stress tensor. 
As we shall see in the following, unlike $\mu_0$, $\mu_f$ does {\em not} diverge as
$\nu\to\nu_c$. In our original model \cite{AT1,AT2} we simply assumed
$\mu_f=const$. 

We postulated \cite{note1} that the fluid part of the 
off-diagonal components of the stress tensor is proportional to the
off-diagonal components of the full stress tensor with the
proportionality  coefficient being a function of the order parameter $\rho$,
\begin{equation}
\sigma^f_{yx}=q(\rho)\sigma_{yx}. 
\label{sigmaf1}
\end{equation}
This assumption stipulates the equation for the static part of the
off-diagonal stress components,
\begin{equation}
\sigma^s_{yx}=(1-q(\rho))\sigma_{yx}.
\label{sigmas}
\end{equation}
Both fluid and solid parts of the stress tensor are assumed symmetric,
$\sigma_{yx}^{f,s}=\sigma_{xy}^{f,s}$. This assumption is confirmed by
our numerical simulations (see below).  We choose a fixed 
range for the order parameter such that it is
zero in a completely fluidized state and one in a completely static
regime. Thus the function $q(\rho)$ has the property $q(0)=1,\ q(1)=0$.
In Refs.\cite{AT1,AT2} for simplicity we took $q(\rho)=1-\rho$. 

A similar relationship can be postulated for the diagonal 
terms of the stress tensor, 
\begin{eqnarray}
\sigma^f_{xx} &=& q_x(\rho)\sigma_{xx},\ \sigma^f_{yy} =
q_y(\rho)\sigma_{yy},
\label{sigmaf2}\\
\sigma^s_{xx} &=& (1-q_x(\rho))\sigma_{xx},\ \sigma^s_{yy} =
(1-q_y(\rho))\sigma_{yy},
\label{sigmas2}
\end{eqnarray}
where the scaling functions $q_{x,y}(\rho)$ can differ from
$q(\rho)$.  

Combining Eqs.(\ref{sigmaf})-(\ref{sigmas2}), we obtain the constitutive
relation in the closed form,
\begin{equation}
\sigma_{\alpha \beta}=p_f\delta_{\alpha \beta}/q_\alpha(\rho)-\mu_f\dot\gamma_{\alpha \beta}/q(\rho)
\label{constit00}
\end{equation}
where $\alpha,\beta\in\{x,y\}$.

The order parameter itself was not related to any microscopic
properties of granular assemblies in Refs. \cite{AT1,AT2}.
We simply assumed that because of strong dissipation in dense granular
flows it has purely relaxational dynamics controlled by the
Ginzburg-Landau equation,
\begin{equation}
\frac{D\rho}{Dt}=D\nabla^2\rho-\frac{\partial {\cal F}(\rho)}{\partial \rho}
\label{GL}
\end{equation}
Here $D$ is the diffusion coefficient and  ${\cal F}(\rho)$ is
the free energy density which was
assumed to have a quartic polynomial form to account for the bistability
near the solid-fluid transition,
\begin{equation}
{\cal F}(\rho)=\int^\rho \rho(\rho-1)(\rho-\delta)d\rho.
\label{G}
\end{equation}
The control parameter $\delta$ is determined by the stress tensor which
in Ref. \cite{AT1,AT2} was taken to be a linear function of $\phi=\max
|\sigma_{mn}/\sigma_{nn}|$, where the maximum is sought over all
possible orthogonal directions $m$ and $n$. It is easy to see that in
the interval $0<\delta<1$ Eq. (\ref{GL}) has two stable uniform solutions
$\rho=0,1$ corresponding to fluid and solid states and one unstable
solution $\rho=\delta$. 

Momentum conservation equation (\ref{momentum}) together with 
Eqs. (\ref{constit00})-(\ref{G}) represent a closed set of continuous
equations which after being augmented by appropriate boundary
conditions, can describe a variety of interesting granular flows
such as avalanches in thin chute flows, drum flows, 
stick-slip oscillation in surface-driven flows, etc. \cite{AT1,AT2}.
Since a rigorous derivation of the continuum model of dense
partially fluidized flows from first principles does not seem feasible, 
the assumptions made {\em ad hoc} in the formulation of the model
(\ref{momentum}),(\ref{constit00})-(\ref{G}) have to be checked
against available experimental and/or numerical data. Unfortunately, it
appears to be extremely difficult to directly measure the order
parameter in the bulk of granular material, as it is determined by
subtle changes in the contact fabric.
In this paper we attempt to extract the properties of the order
parameter from molecular dynamics simulations and on this basis make a
quantitative fit of the order parameter model.

\section{Molecular Dynamics Simulations \label{sec:mds}}

To model the interaction of individual grains we use the so-called soft-contact
approach. The grains are assumed to be non-cohesive, dry, inelastic
disk-like particles. Two
grains interact via normal and shear forces whenever they overlap.  For
the normal impact we employ \emph{spring-dashpot} model \cite{wolf96}.
This model accounts for repulsion and dissipation; the repulsive
component is proportional to the degree of the overlap, and the velocity
dependent damping component simulates the dissipation.  The model for
shear force is based upon the technique developed by Cundall and Strack
\cite{cundall79}.  It incorporates tangential elasticity and Coulomb
laws of friction. The elastic restoring force is proportional to the
integrated tangential displacement during the contact and limited by the
product of the friction coefficient and the instantaneous normal force.
The grains possess two translational and one rotational degrees of
freedom.  The motion of a grain is obtained by integrating the Newton's
equations with the forces and torques produced by its interactions with
all the neighboring grains and walls of the container.

Consider a grain $i$ of radius $R^i$ located at ${\bf r}^i$ moving with
translational velocity ${\bf v}^i$ and angular velocity $\omega^i$. This
grain is in contact with grain $j$ whenever overlap $\delta_n
= R^i + R^j - |{\bf r}^i-{\bf r}^i| > 0$. The relative velocity at the
contact point and its normal and tangent components are given by
\begin{eqnarray}
\label{vc}
{\bf v}^{ij} &=& {\bf v}^i - {\bf v}^j  + (R^i\omega^i + R^j\omega^j) {\bf t}^{ij},\\
\label{vcnt} 
v_n &=& ({\bf v}^i - {\bf v}^j ){\bf n}^{ij}, \quad v_t = ({\bf v}^i - {\bf v}^j ){\bf n}^{ij} + (R^i\omega^i + R^j\omega^j),
\end{eqnarray}
where ${\bf n}^{ij} = ({\bf r}^i -{\bf
r}^i)/r^{ij}=(n_x,n_y)$ is the inward normal to the surface of $i$ at
the contact point with $j$, and the direction of the tangent ${\bf t}^{ij} = (n_y,
-n_x)$ is chosen so that ${\bf t}^{ij}\times {\bf n}^{ij}$ is co-linear
with the angular velocity.  Then the grain $i$ is subjected to the contact force
due to the interaction with $j$ with normal and tangential components
\begin{eqnarray}
\label{fn}
F_n &=& k_n \delta_n - 2 \gamma_n m_e v_n,\\
\label{ft}
F_t &=& - \mbox{sign}(\delta_t) \min (|k_t \delta_t|, |\mu F_n|), \quad \delta_t(t) = \int_{t_0}^{t}{v_t(\tau)}d\tau,
\end{eqnarray}
where $k_{n,t}$ are the corresponding spring constants, $\gamma_n$ is
the normal
damping coefficient, $m_e=m^i m^j/(m^i+m^j)$ is the reduced mass, $\mu$ is
coefficient of friction, and $\delta_t$ is tangential displacement since
the moment $t_0$ of the initial contact between $i$ and $j$. When the
static yield criterion Eq. (\ref{ft}) is satisfied, the magnitude of
$\delta_t$ is adjusted to an instantaneous equilibrium value providing
$F_t = \mu F_n$.  According to the analytical solution of the linear
spring-dashpot model the coefficient of restitution and the duration of
heads-on collision are $e =\exp(-\gamma_n t_n)$, $t_n = \pi/(k_n/m_e
- \gamma_n^2)^{1/2}$.

The advantages and limitations of the employed contact force model were
thoroughly studied by a number of authors \cite{wolf96,grest91,grest02}. 
In fact this is the simplest model which allows us to account for both static
and dynamic friction.

When all forces acting on grain $i$ from other grains, boundaries and
perhaps external fields are computed, the problem is reduced to the
integration of the Newton's equations for translational and rotational
degrees of freedom,
\begin{eqnarray}
\label{newttrans}
m^i \frac{d^2 {\bf r}^i}{dt^2} &=& m^i {\bf g} + \sum_{c}{{\bf F}^{ic}},\\
\label{newtrot}
I^i \frac{d \omega^i}{dt} &=& R^i \sum_{c}{F_t^{ic}},
\end{eqnarray}
where the mass of particle $i$ is denoted by $m^i$ and its moment of inertia is $I^i=1/2 m^i {R^i}^2$,
$m^i{\bf g}$ stands for an external gravity field, and the sums in Eqs. (\ref{newttrans}),(\ref{newtrot})
run over all contacts of particle $i$. 

The results of the simulations reported here are presented in
dimensionless form.  All quantities are normalized by an appropriate
combination of the average particle diameter $d$, mass $m$, and gravity $g$.

Eqs. (\ref{newttrans}),(\ref{newtrot}) were integrated using the
fifth-order predictor-corrector \cite{allen87} with a constant time step
$\delta  t$.
The spring constant $k_n$  and damping coefficient $\gamma_n$ were
chosen to provide the desired value of the restitution coefficient $e$
and  guarantee an accurate resolution of an individual collision.
Typically, we used $e=0.92$, $\delta t = 10^{-4}$, $t_c=50 \delta t$,
$k_n=2.0\ 10^5$, $\gamma_n = 16.7$, $k_s=1/3 k_n$.

The computational domain spans $L_x\times L_y$ area, and is periodic in
horizontal direction $x$.  Unless otherwise mentioned the granulate is
slightly polydisperse to avoid crystallization effects \cite{grest91}.
We assume that the grain diameters are uniformly distributed around mean
with relative width $\Delta_r$.
To provide a link between micro-mechanical quantities obtained through
simulations and continuous fields, we define the following coarse-graining
procedure. Since all experiments described below deal with
steady quantities, the procedure consists of two steps: space and then
time averaging.  The space averaging of a field $\emph{x}$ is performed
over horizontal bins (along the flow direction) of size $V_b\approx
L_x\times d$ and is denoted as $\langle \emph{x}\rangle$ in the
following.  Contributions of those particles which only partially
belong to a certain bin are weighted by the fraction of their area.
 After a simulation has reached a steady state, 
instantaneous profiles are averaged over a suitable number of
time-snapshots. We shall denote the time-averaging  with
$\overline{\emph{x}}$.

For example, steady solid fraction profile is given by
\begin{equation}
 \nu(y) = \overline{\langle\nu(y,t)\rangle}, \quad
\langle\nu(y,t)\rangle=V_b^{-1}\sum_{i\in V_b} w^i(y)V^i,
\label{nu}
\end{equation}
where the summation runs over grains at least partially in a bin $V_b$
centered at $y$, $V^i$ is the area of grain $i$, and $w^i(y)$ is a 
corresponding fraction of a grain's
area within $V_b$. The coarse-grained velocity field is
\begin{equation}
{\bf u}(y) = \overline{\langle{\bf u}(y,t)\rangle},\quad 
\langle {\bf u}(y,t)\rangle = \nu(y,t)^{-1}\langle{\bf v}^i(y,t)\rangle =
\nu(y,t)^{-1} V_b^{-1}\sum_{i\in V_b} w^i(y) V^i{\bf v}^i.
\label{vel}
\end{equation}
For simplicity we extend the space-time averaging technique described
above for other quantities, such as the stress tensor
\begin{equation}
\sigma_{\alpha\beta}(y) = \overline{\langle \sigma_{\alpha\beta}(y,t)\rangle},\quad
\langle\sigma_{\alpha\beta}(y,t)\rangle =
\left \langle \frac{1}{2}\sum_{c\ne i}r^{ic}_{\alpha} F^{ic}_{\beta} \right \rangle + 
\langle m^i \tilde{v}^i_{\alpha} \tilde{v}^i_{\beta} \rangle,
\label{sigma}
\end{equation}
where $\alpha,\beta =\{x,y\}$, $r^{ic}_{\alpha} = {\bf r}^{ic}\cdot {\bf e}_{\alpha}$,
$F^{ic}_{\beta} ={\bf F}^{ic}{\bf e}_{\beta}$, 
$\tilde{v}^i_{\alpha} = v^i_{\alpha} - \overline{\langle v^i_{\alpha}\rangle}$.
The stress tensor  in Eq. (\ref{sigma}) has two distinct components. The
first one -- virial or contact -- describes pairwise interactions of grains.
The second one -- kinetic or Reynolds -- is due to velocity fluctuations.

\subsection{Order parameter for granular fluidization:
static contacts vs. fluid contacts}

At any moment of time all contacts are classified as either
``fluid'' or ``solid''.  A contact is considered ``solid'' if
it is in a stuck state ($F_t < \mu F_n$) and its duration is longer
than a typical time of collision $t_*$. The first requirements
eliminates long-lasting sliding contacts, and the second requirement 
excludes short-term collisions pertinent for completely fluidized
regimes.  We choose a typical collision to last $t^*=1.1 t_c$. When either
of the requirements is not fulfilled, the contact is assumed ``fluid''. 

We define the order parameter as the ratio between 
space-time averaged numbers of ``solid'' contacts $\overline{\langle
Z_s\rangle}$ and 
all contacts $\overline{\langle Z\rangle}$ within a sampling area
\cite{silbert02},
\begin{equation}
\rho(y)=\overline{\langle Z_{s}^i\rangle} / \overline{\langle Z^i\rangle} .
\label{rho}
\end{equation}
This definition endures at least two limiting cases: when a
granulate in a static state and when it is strongly agitated, i.e.
completely fluidized. In fact, in the former quiescent state all
contacts are stuck and $\rho = 1$. In the fluidized case
$\overline{\langle Z_s\rangle}$ is zero and 
$\overline{\langle Z\rangle}$ is small but finite, therefore $\rho = 0$. 

Let us note that the order parameter $\rho$ just introduced is expected
to be as sensitive to the degree of fluidization as the stress tensor.
A small rearrangement of the force network may result in
strong fluctuations of either field, while such quantities as the solid
fraction or granular temperature will remain virtually constant. It is 
known that granular
aggregates exhibit rigidity phase transition near the critical volume
fraction $\nu_c$ \cite{sparks99}. When the volume fraction is near or
above $\nu_c$,  the granular material has elasto-plastic rheology and
below this critical value it behaves as a fluid.  Near and above the 
critical volume fraction for the same boundary conditions the
granular aggregate may exhibit either slow creeping flow, solid-like
behavior, or both. The latter case, known as stick-slip phenomenon, is
observed both experimentally and numerically
\cite{behringer96,gollub98,sparks99}. In an experiment under constant
volume boundary conditions a series of stick-slip events occur without
a significant change in the solid fraction, which is insensitive to global 
rearrangements relieving the accumulated stress. We expect the order parameter to be
able to reflect such rearrangements and ultimately describe 
the corresponding phase transition. 

\subsection{Stress tensor}

The full stress tensor (\ref{sigma}) consists of a contact (virial) part and the 
Reynolds part. In turn, the contact stress tensor can be
split into the ``solid contact'' component, $\sigma^{s}$, and
the ``fluid contact'' component, ($\sigma^{f}$), in the same fashion as was
done with contacts themselves. Combining the ``fluid contact'' component
with the Reynolds stress, we obtain the full stress tensor as a sum of
two parts
\begin{eqnarray}
\label{stress_decomposition}
&&\sigma=\sigma^{f}+\sigma^{s},\\
\label{stresses}
&&\sigma^{s}_{\alpha\beta} = \left \langle \frac{1}{2}\sum_c^{\prime} r^{c}_{\alpha} F^{c}_{\beta} \right \rangle,\quad
\sigma^{f}_{\alpha\beta} = \left \langle \frac{1}{2}\sum^{\prime \prime} _{c}r^{c}_{\alpha} F^{c}_{\beta} \right 
\rangle + \langle m^i \tilde{v}_{\alpha} \tilde{v}_{\beta} \rangle
\end{eqnarray}
where summation in $\sum^\prime $ and $\sum^{\prime \prime} $ is restricted to 
``solid'' and ``fluid'' contacts respectively.  

The ``fluid'' part of the stress tensor is due to short-term collisional
stresses and the Reynolds stresses, whereas the solid part accounts for 
persistent force chains. The
Reynolds contribution to the stress is negligibly small in the
vicinity of the phase transition, but comes into play when the granular
aggregate is highly fluidized.  In the system which is neither
completely rigid nor completely fluidized we expect the coexistence (in
time and space) of both phases. A particular grain may have both
types of contacts at the same time thus contributing to both
$\sigma^{f}$ and $\sigma^{s}$. This picture is reminiscent of
the concept of bimodal character of stress transmission in static contact
network introduced by Radjai et al \cite{moreau96,moreau98}. 

\section{Testbed system: Couette flow in a thin granular layer
\label{sec:testbed}}
\subsection{Bifurcation diagram}
\begin{figure}[ptb]
\includegraphics[width=3.5in]{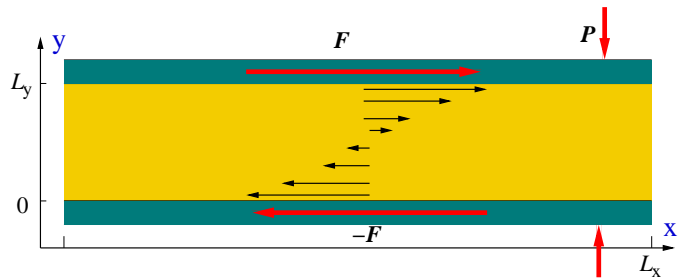}
\caption{Sketch of the granular shear flow model}
\label{setup_shear}
\end{figure}

\begin{figure}[ptb]
\includegraphics[width=3.5in]{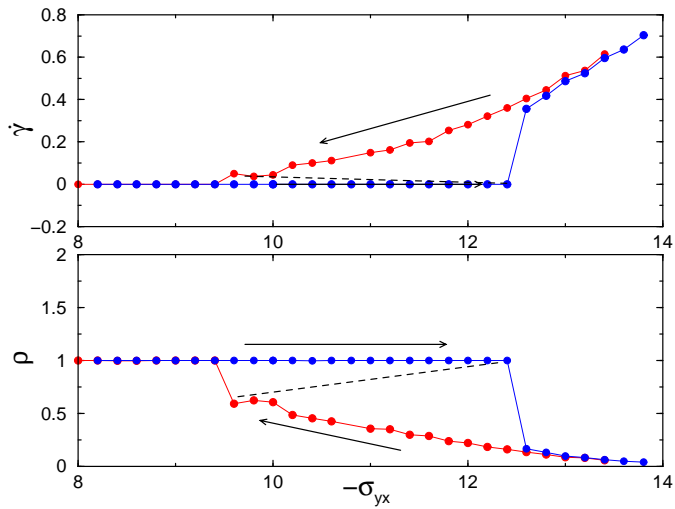}
\caption{The strain rate (a) and the order parameter (b) 
vs. shear stress in a thin Couette geometry of
Figure \protect\ref{setup_shear} with 500 particles (10x50) at  $P=40$.
}
\label{transition}
\end{figure}

We studied the fluidization transition in a thin ($50\times 10$) granular layer
between two ``rough plates'' under fixed pressure $P$ and zero gravity conditions
(Figure\ref{setup_shear}). 
The parameters of simulations were $\mu_0=0.5,\ e=0.92,\  \Delta_r=0.2,\
k_n=2\cdot10^{5},\ k_s/k_n=1/3,\ \gamma_n=16.7$.
We used the same material parameters of grains throughout this Section.
The rough plates were simulated by two straight
chains of large grains (twice as large as an average particle diameter) glued 
together. The layer was chosen narrow enough, so the properties of
the granular layer (shear rate, shear stress components, order parameter,
etc) were constant across the layer. Two opposite forces
${\bf F}_1=-{\bf F}_2$ were applied to the plates along the horizontal 
$x$ axis to induce shear stress
in the bulk.  We started with weak forces not sufficient to initiate a
shear flow and slowly ramped them up in small increments well above the
critical yield force at which the granular flow started. After that we ramped
the shear forces down until the granular layer was jammed again. At
every ``stop'' we measured all stress components, strain rate, and the
order parameter, and averaged the data over the whole layer and over
time of each step. 
Figure
\ref{transition} shows the strain rate $\dot\gamma$ (we drop subscripts
${yx}$ at the strain rate) and the order parameter
$\rho$ as functions of the shear stress $\sigma_{yx}$ which is
approximated by the applied force $F$ normalized by the layer length
$L_x$, for $P=40$.  
As it is to be expected, the strain rate remains zero, and the
order parameter is one until the shear stress reaches a certain critical
value $\sigma_1\approx 12.6$. This value differs slightly for different
runs because of the finite system size and absence of self-averaging in
the static regime. Above the yield stress, the strain rate abruptly
jumps to a finite value $\approx 0.35$, and the order parameter drops to
$\approx 0.15$. At larger $|\sigma_{yx}|$, the strain rate increases
faster than linearly, and the order parameter rapidly approaches zero.
The return curve corresponding to the diminishing of the shear stress
follows roughly the same path, and then continues to another (smaller)
value of the shear stress ($\sigma_2\approx 9.4$).  At this value the layer
jams, the strain rate returns to zero, and the order parameter
jumps back to one. 

\begin{figure}[ptb]
\includegraphics[width=3.5in]{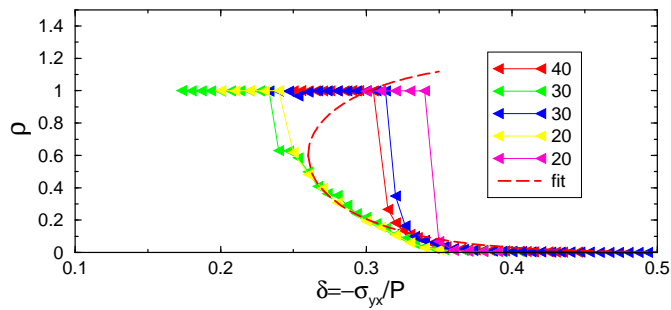}
\caption{The order parameter as a function of
the normalized shear stress in a thin Couette geometry. 
}
\label{norm}
\end{figure}

The most striking feature of this figure is the hysteretic behavior of
both the strain rate and the order parameter as a function of the shear
stress. This hysteresis was anticipated in our order-parameter model
\cite{AT1,AT2}, however now we are in a position to fit the model equations
quantitatively.  We repeated these 
simulations at several different values of the compressing pressure $P$.
Data for different pressure values in the flow regime fall onto the same
universal curve if one normalizes the shear stress by the pressure (see
Figure \ref{norm}). Assuming that there is an (unobserved) unstable
branch of the bifurcation diagram which merges with the stable branch
at $\delta\approx 0.26$, we can make a simple analytic fit  of this curve
as
\begin{equation}
F(\rho,\delta)=(1-\rho)\left(\rho^2-2\rho_*\rho+\rho_*^2\exp[-A(\delta^2-\delta_*^2)]\right)=0 
\label{FF}
\end{equation}
with $\rho_*=0.6, A=25,\delta_*=0.26$
(see Figure \ref{norm}, line) and use it in the polynomial expansion of the
free energy density which enters the order parameter equation
(\ref{GL}): 
\begin{equation}
{\cal F}(\rho)=\int^\rho F(\rho,\delta)d\rho.
\label{F1}
\end{equation}

\begin{figure}[ptb]
\includegraphics[width=2.5in]{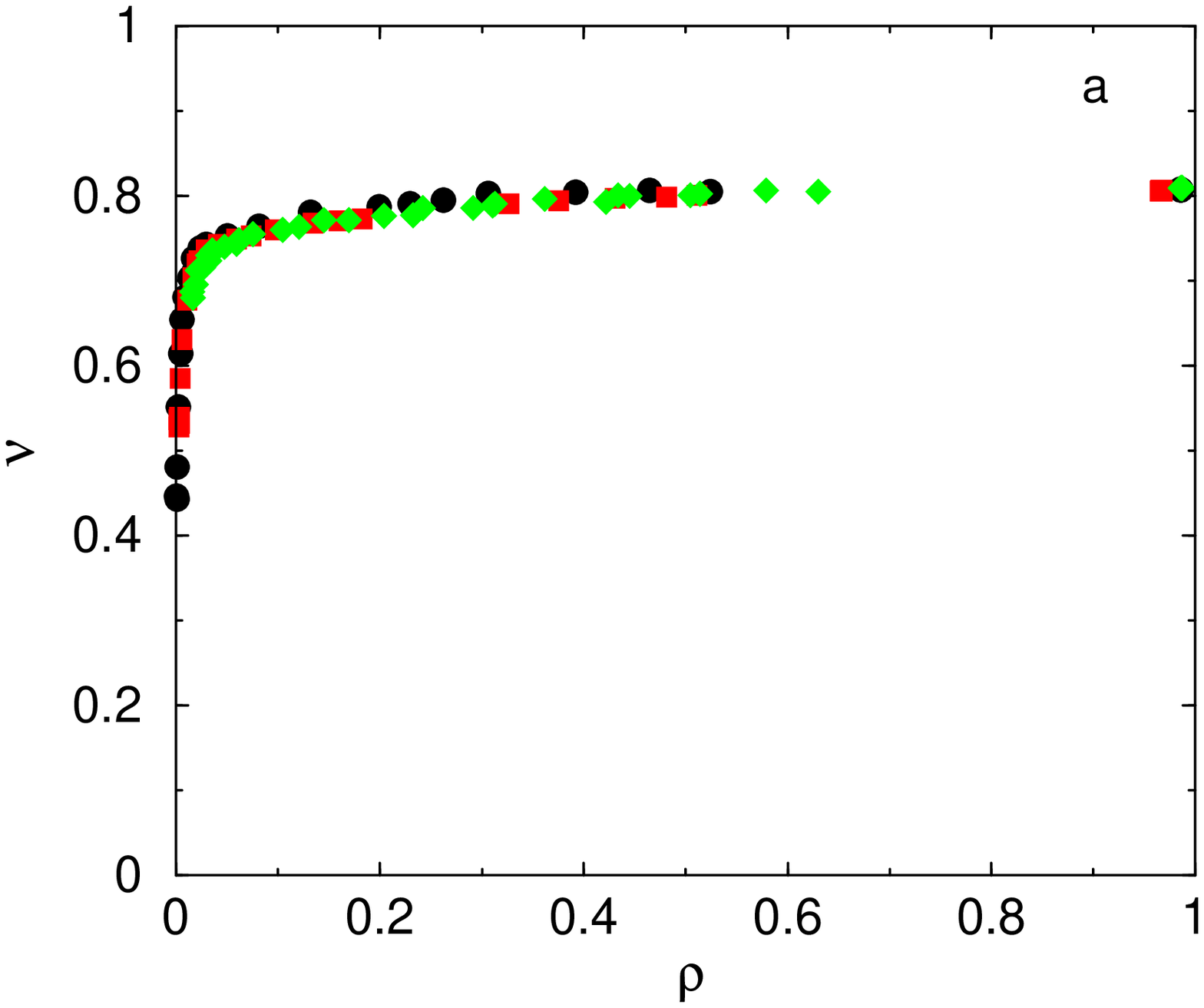}
\includegraphics[width=2.5in]{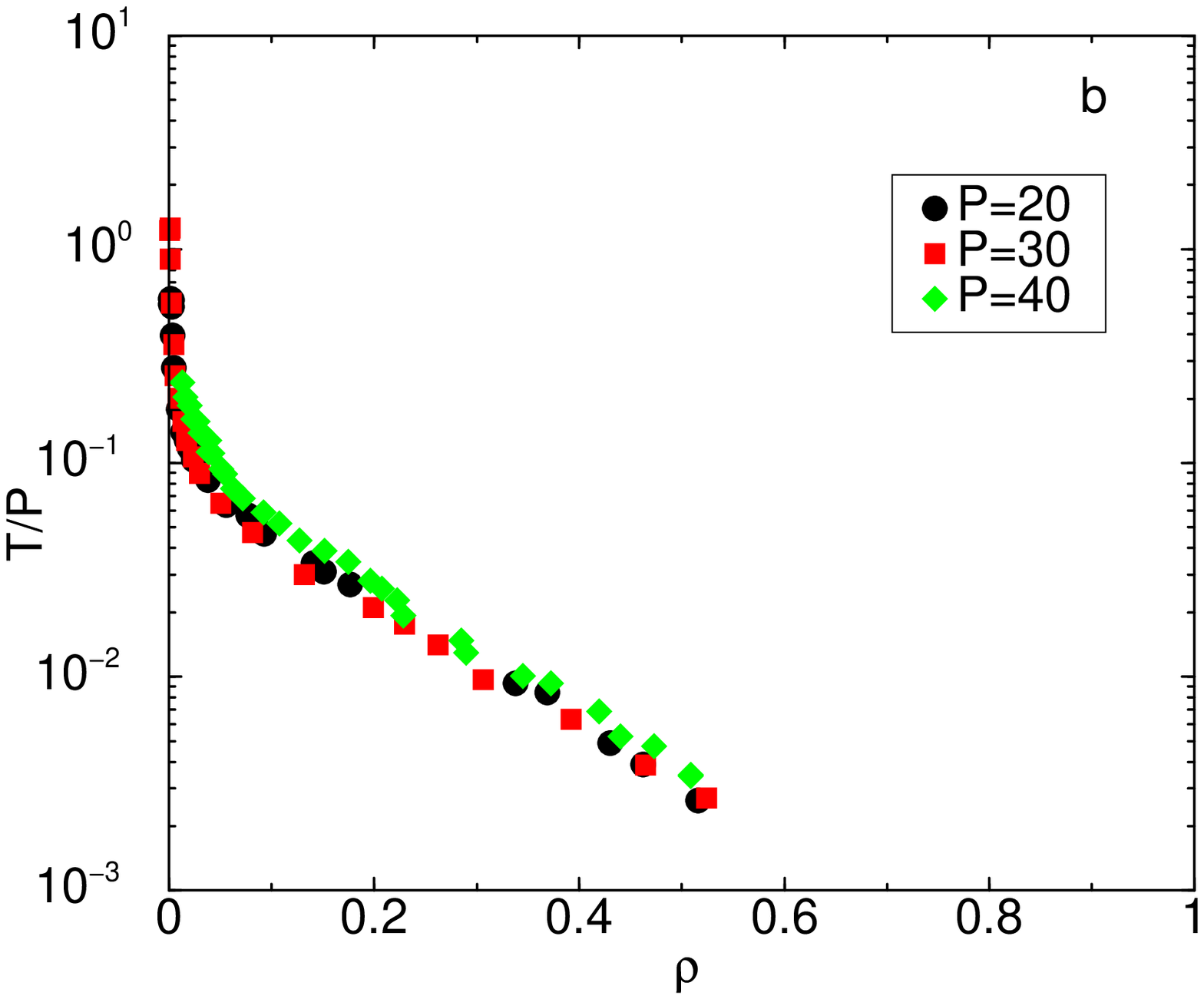}
\caption{The density (a) and the normalized granular temperature (b) 
versus the order parameter in a thin Couette geometry for three different 
values of pressure $P$.
}
\label{op_T_nu}
\end{figure}

We also measured the density and the granular temperature of the grains
as we decreased the shear force. These measurements could only be
performed in the range $\rho<0.55$ since for larger $\rho$ the partially
fluidized state is unstable.  Note that that density of grains stays
almost constant in a wide range of the order parameter $\rho>0.1$ (see
Figure \ref{op_T_nu}a).  The granular temperature (which is defined as
$T=\langle \tilde{u}_i\tilde{u_i}\rangle/2$) normalized by the applied
pressure $P$ appears to be a unique function of the order parameter
(Figure \ref{op_T_nu}b). 

\subsection{Relaxation dynamics of the order parameter}
\begin{figure}[ptb]
\includegraphics[width=3.5in]{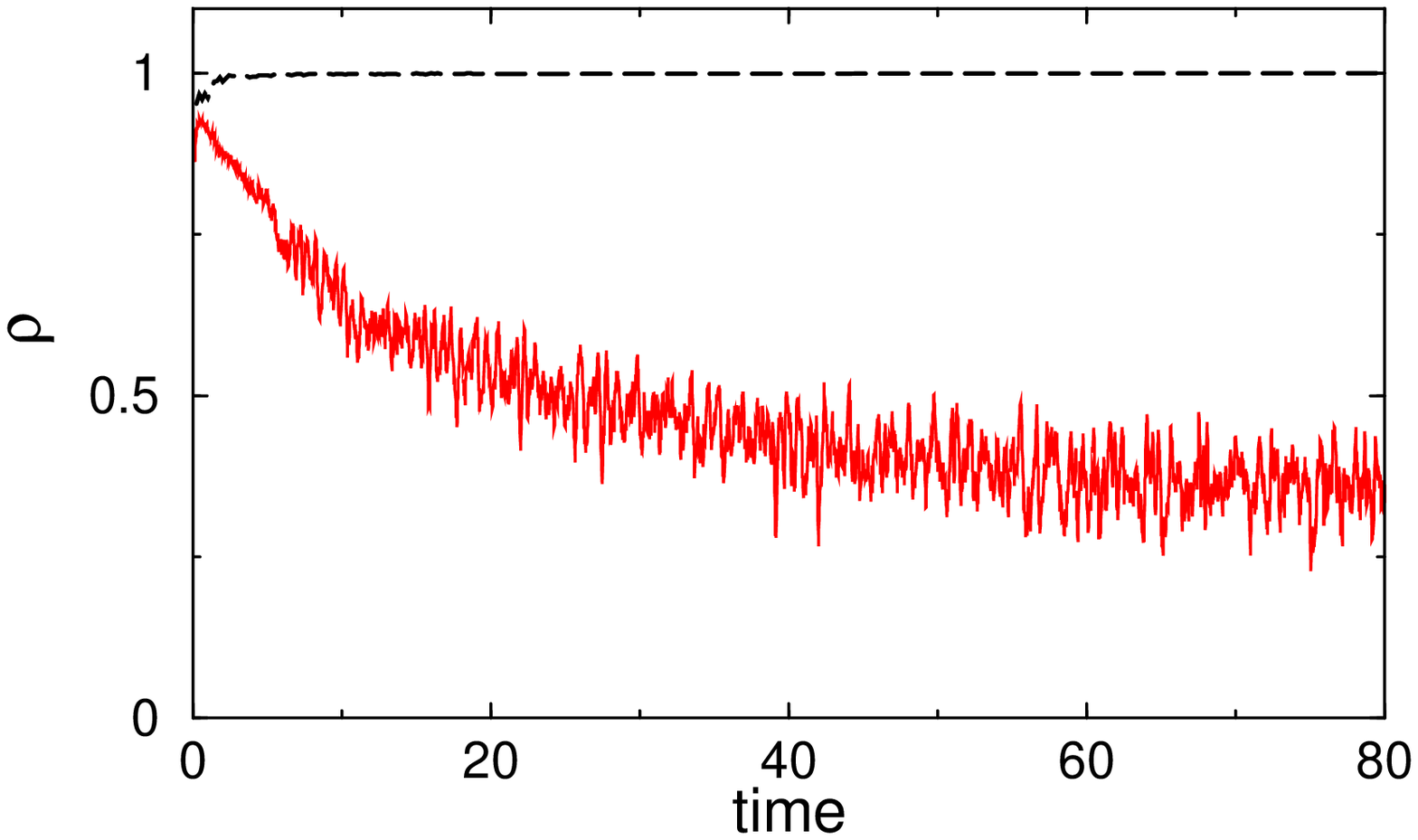}
\caption{Relaxation of the order parameter towards shear flow 
(solid line) and jammed state (dashed line) for two different initial
perturbations in a thin Couette system of 500 particles with 
$P=40, \sigma_{yx}=-12$.
}
\label{rho_relax}
\end{figure}

We probed the relaxation dynamics of the order parameter by performing
the following numerical experiment. The granular layer was prepared as in
the previous Section. Lateral shear forces were increased adiabatically
until the granular system reached a metastable solid (jammed) state within a hysteretic 
region. Then the layer was perturbed by applying random forces to a small randomly 
selected fraction of particles. The
dynamics of the order parameter vary depending on the magnitude of the
perturbation. Figure \ref{rho_relax} shows an example of the evolution
from the same jammed state for two different magnitudes of initial
perturbation. Interestingly, the relaxation back to the jammed state is
very fast, whereas the relaxation toward stable shear flow is
much slower. We believe that it has to do with inertia of grains and
the upper plate, so the intrinsic time scale of the order parameter
relaxation is rather small, $O(1)$.

Unfortunately, our thin Couette flow system does not allow us to probe
the local coupling of the order parameter since the order parameter is
uniformly distributed throughout the system.  In the absence of this
data, this coupling was
modeled by the linear diffusion term in (\ref{GL}) with the constant
diffusion coefficient. As we will see in the following Section, this
approximation indeed provides a good description for spatially {\em
non-uniform} near-surface flow, however the value of the diffusion
coefficient appears to be a function of local stress.

\subsection{Fitting the constitutive relation}

\begin{figure}[ptb]
\includegraphics[angle=0,width=3.in]{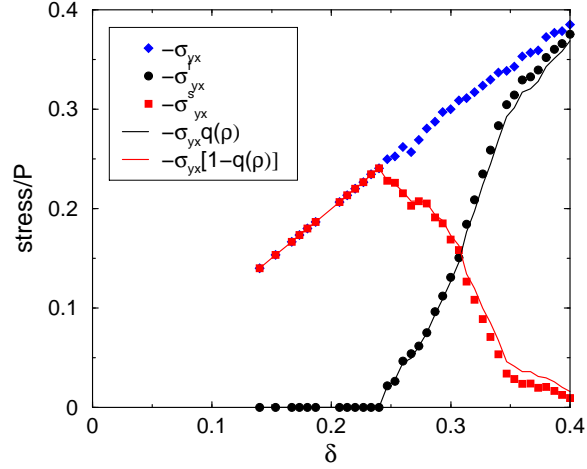}
\vskip 0.5cm
\caption{Fluid and static components of the shear stress in a thin granular layer (10x50)
at external pressure $P=30$ as a function of normalized external shear stress
$\delta=F/(L_xP)$:
direct calculation (points), obtained from total stress using
relations (\protect\ref{sigmaf1}),(\protect\ref{sigmas}) 
with $q(\rho)=(1-\rho)^{2.5}$ (lines).
}
\label{stress30}
\end{figure}

\begin{figure}[ptb]
\includegraphics[angle=0,width=2.5in]{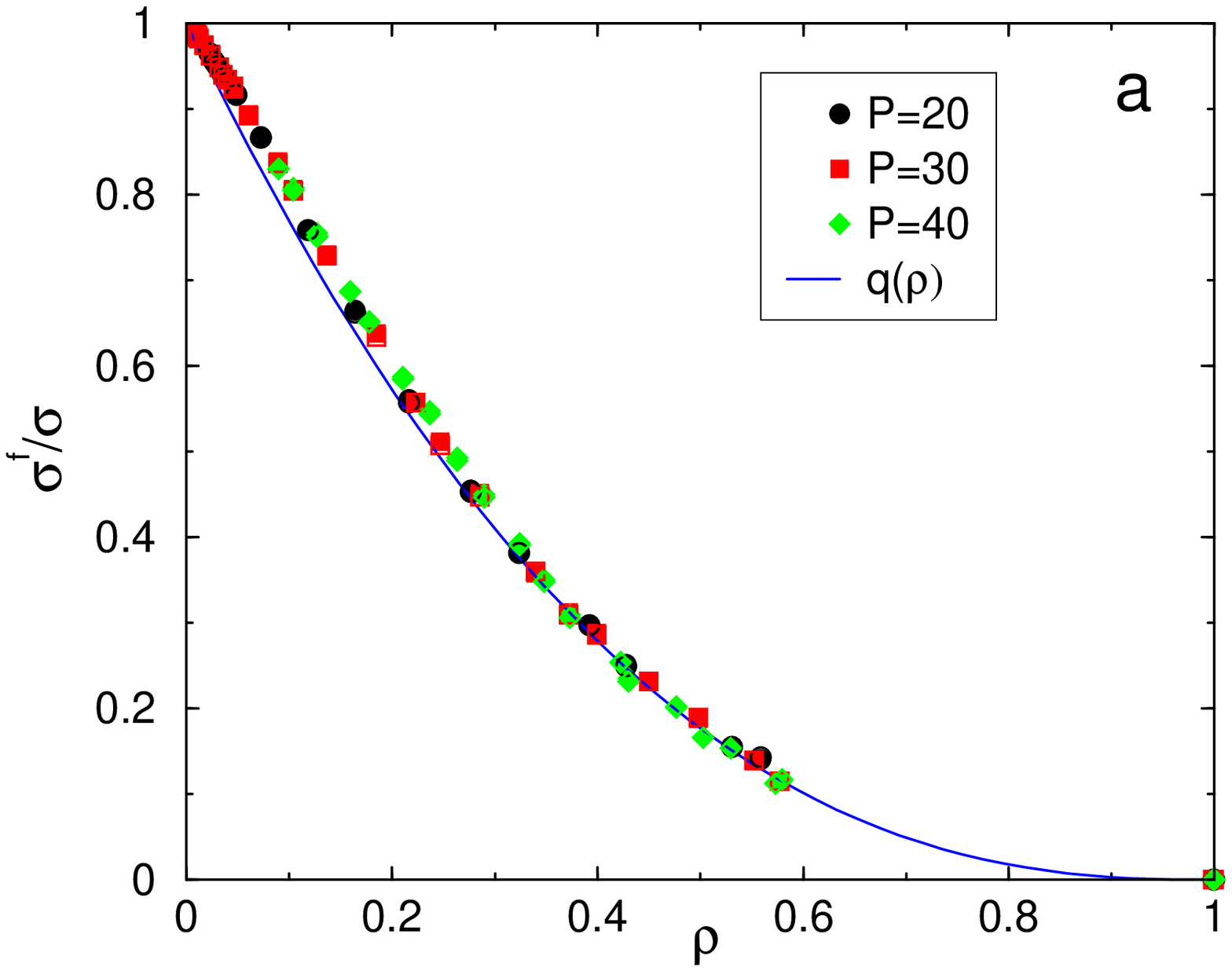}
\includegraphics[angle=0,width=2.5in]{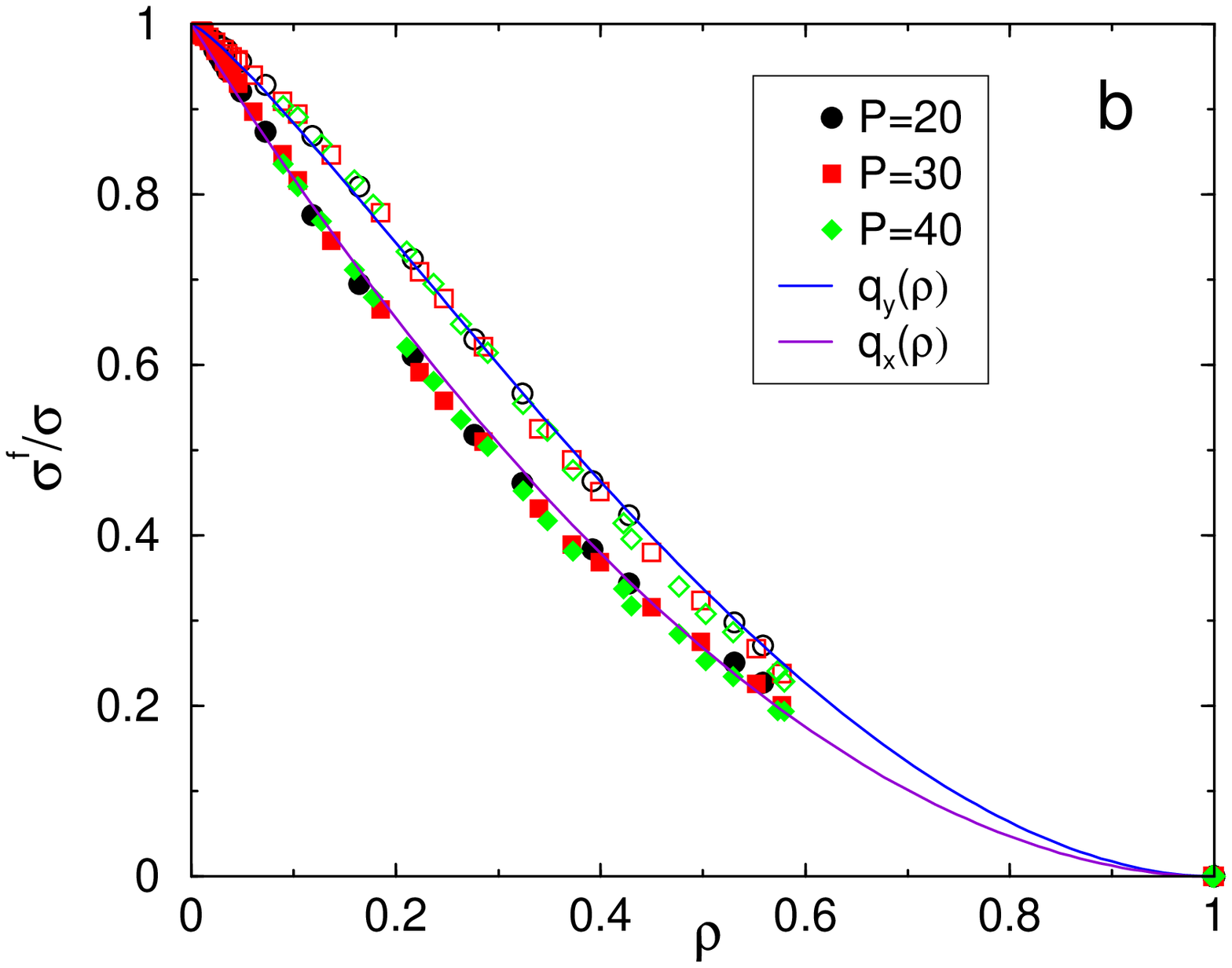}
\vskip 0.5cm
\caption{Ratios of the fluid stress components to the corresponding full 
stress components
$\sigma^{f}_{\alpha \beta}/\sigma_{\alpha \beta}$ vs. $\rho$ for three
different pressures $P$: a - shear stress components, closed symbols
$\sigma_{yx}$, open symbols $\sigma_{xy}$, line is a fit 
$q(\rho)=(1-\rho)^{2.5}$.
b - normal stress components, closed symbols $\sigma_{xx}$, open symbols
$\sigma_{yy}$, lines are the fits $q_x(\rho)=(1-\rho)^{1.9},\
q_y(\rho)=(1-\rho^{1.2})^{1.9}$. 
}
\label{constit}
\end{figure}

The next step is to fit the constitutive relation from MD simulations.
To this end, we use the same Couette flow simulations, but now
we analyze the ``fluid stress" $\sigma^f_{\alpha \beta}$ and the ``static 
stress" $\sigma^s_{\alpha \beta}$ separately during our ramp-down simulations at
three different values of $P$. Figure \ref{stress30} shows a sample of these data for
$P=30$. At large $\delta=F/(L_xP)$, when the order parameter $\rho$ is low,  
the total stress is dominated by the fluid component, but as the flow
stops and $\rho$ approaches unity, the fluid stress turns to zero, and the
total stress is equal to the static stress. Plotting
$\sigma^f_{yx}/\sigma_{yx}$ as a function of the
order parameter $\rho$ for different $P$
(Figure \ref{constit}a), we observe that all data collapse onto a
single curve which is well fitted by $q(\rho)=(1-\rho)^{2.5}$.
The lines in Figure \ref{stress30} show the fit of the fluid and static stress
tensors using (\ref{sigmaf1}),(\ref{sigmas}) with
$q(\rho)=(1-\rho)^{2.5}$.  

\begin{figure}[ptb]
\includegraphics[angle=0,width=3.in]{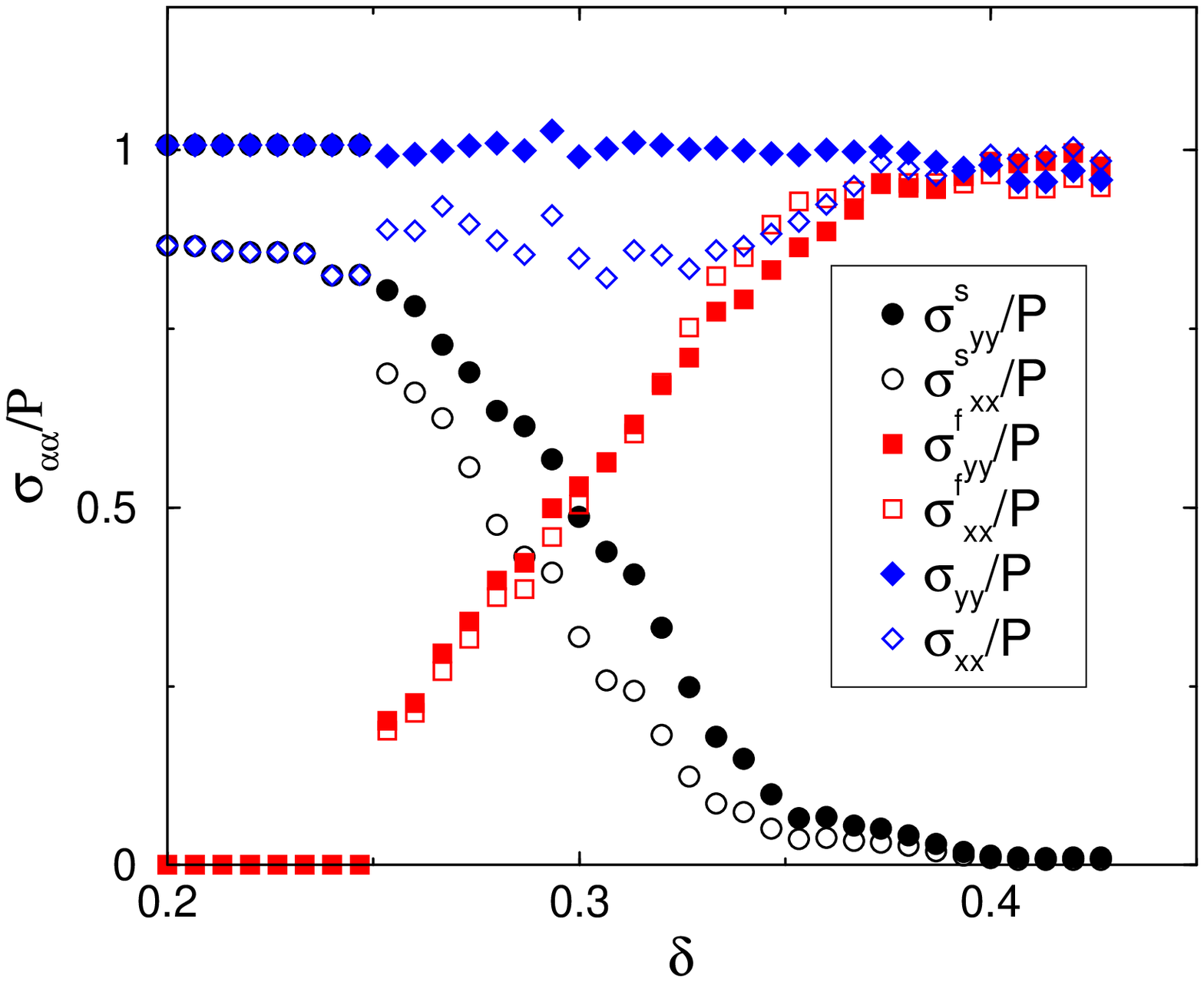}
\vskip 0.5cm
\caption{Diagonal stress components
(fluid, solid, and total) in a thin granular layer (10x50)
at external pressure $P=30$ as a function of external shear stress
$F/L_x$ normalized by the external pressure $P$.
}
\label{pressure}
\end{figure}

The fluid as well as solid parts of the stress tensor are nearly
symmetric, $\sigma^{f,s}_{yx}=\sigma^{f,s}_{xy}$, so the ratio
$\sigma^f_{xy}/\sigma_{xy}$ is described by the same scaling function
$q(\rho)$.  On the other hand, the same procedure for the diagonal
elements of the stress tensor yields a noticeably different scaling (see
Figure\ref{constit},b). Furthermore, a small but noticeable difference is
evident between $\sigma^f_{xx}/\sigma_{xx}$ and $\sigma^f_{yy}/\sigma_{yy}$. 
More detailed analysis shows that in fact fluid parts of the diagonal
components of the stress tensor $\sigma^f_{xx}$ and $\sigma^f_{yy}$ are 
nearly identical, and the difference is due to the solid part of the 
normal stresses (see Figure \ref{pressure}). 
This observation is consistent with the fact that
the diagonal terms of the static stress tensor are determined by the details of
the external loading. On the other hand, in a completely fluidized state
the diagonal terms are all equal to the hydrodynamic pressure
$p_f$ \cite{walton86}. In a
partially fluidized regime, the diagonal terms of the shear stress 
can be expressed as
\begin{equation}
\sigma_{xx}=p_f/q_x(\rho),\ \sigma_{yy}=p_f/q_y(\rho).
\label{constit2}
\end{equation}
Both functions $q_{x,y}(\rho)$ should approach 1 as $\rho\to 0$, but
they may have different functional form to reflect the
anisotropy of the static stress tensor. If the fluid pressure $p_f$
approaches zero in the solid state as $p_f\propto (1-\rho)^\alpha$, then $q_{x,y}\propto
(1-\rho^{\beta_{x,y}})^\alpha$, so $\sigma_{xx,yy}\propto\beta_{x,y}^\alpha$. 
In our simple Couette flow, the diagonal stress tensor components can
be well fitted by $q_x(\rho)\approx (1-\rho)^{1.9}$ and
$q_y(\rho)\approx (1-\rho^{1.2})^{1.9}$ (see Figure \ref{constit},b). 
So we observe that even
in a partially fluidized regime, the ``fluid phase'' component indeed 
behaves as a real fluid with a well-behaved ``partial'' pressure $p_f$
which is zero in a solid state at $\rho=1$ and is becoming the full 
pressure in a completely fluidized state $\rho=0$.

\begin{figure}[ptb]
\includegraphics[angle=0,width=2.5in]{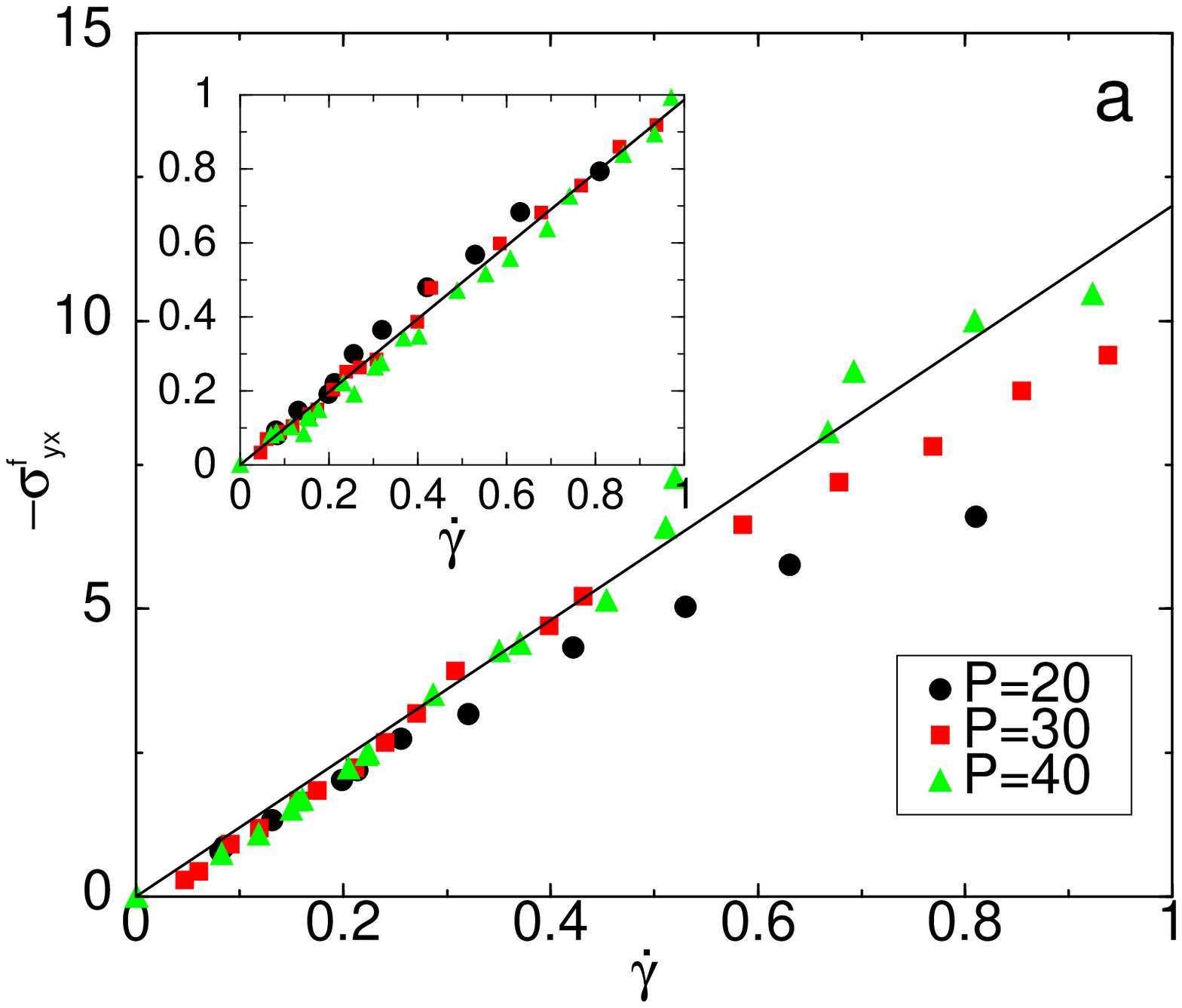}
\includegraphics[angle=0,width=2.5in]{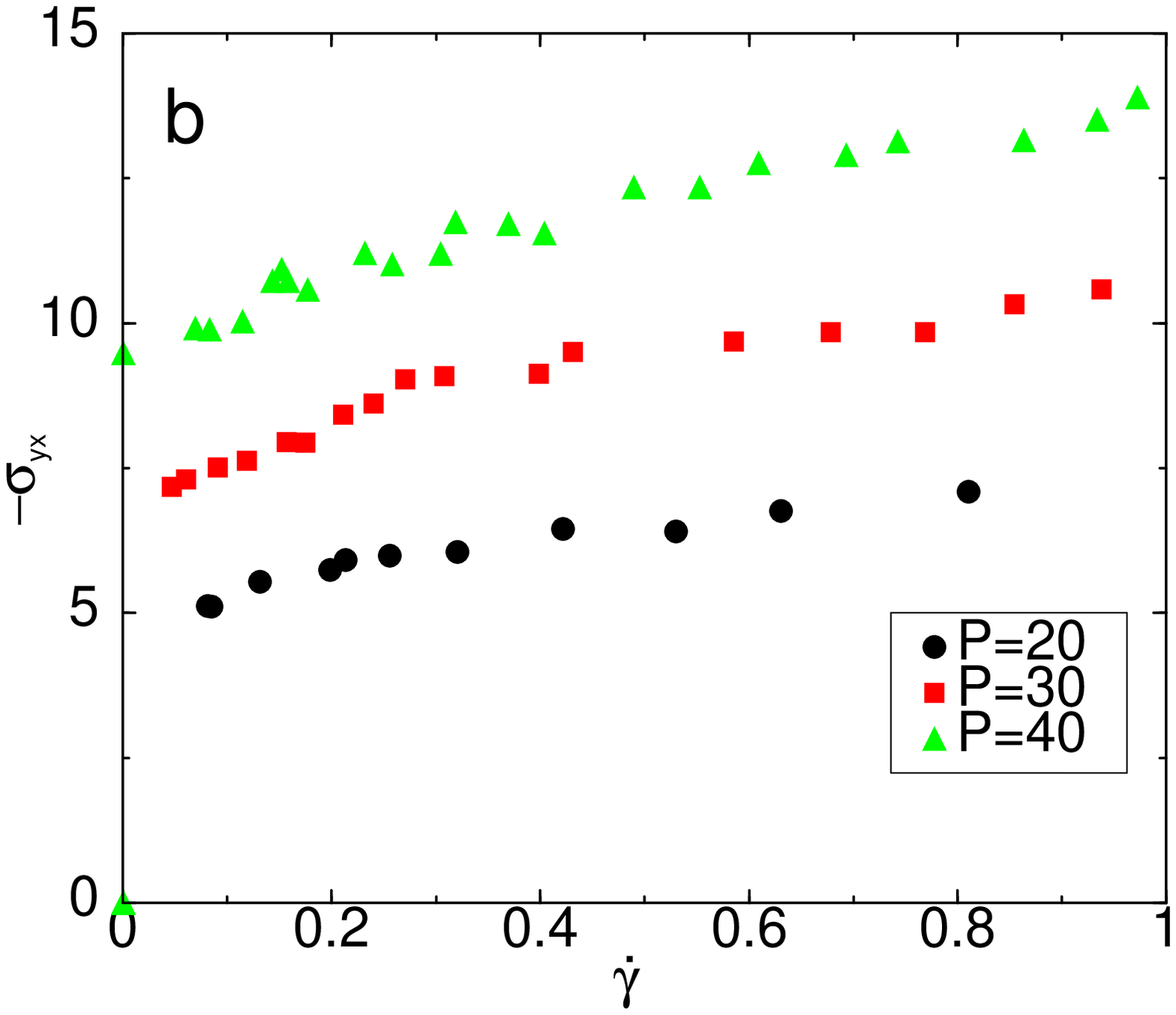}
\vskip 0.5cm
\caption{Stress-strain rate relation for a thin granular Couette flow at
three different external pressures: {\em (a)} fluid shear stress vs strain
rate, the straight line is a constant viscosity fit
$\sigma^f=12\dot\gamma$; {\em inset}: scaled fluid shear stress
$\mu_f^{-1}\sigma^f_{yx}$ as a function of $\dot\gamma$;
{\em (b)} full shear stress vs strain rate
}
\label{visc}
\end{figure}

Plotting the fluid shear stress versus the strain rate, we can test the
validity of the 
Newtonian model for the stress-strain relation (\ref{sigmaf}). Figure
\ref{visc} shows $-\sigma^f_{yx}$ vs $\dot\gamma$ for three different
pressures $P=20,30,40$. At small $\dot\gamma$ all three lines are close to
the same straight line $\sigma^f_{yx}=12\dot\gamma$,
which indicates that the Newtonian scaling for fluid shear stress holds
reasonably well. The deviations at large $\dot\gamma$ are evidently caused by
variations of temperature and density in the dilute regime. 
Note that in contrast, the full shear stress does not goes to zero as
$\dot\gamma\to 0$ (see Figure \ref{visc}b), so a viscosity coefficient
conventionally defined as the ratio of the full shear stress to the
strain rate diverges at the fluidization threshold as observed in
Ref. \cite{gollub00}.

Combining the Newtonian law for the fluid stress-strain dependence with 
the order parameter scaling of the fluid stress tensor, we arrive at the 
relationship between the full stress tensor and the strain rate tensor 
(\ref{constit00}) with $\mu=12,\ q_x(\rho)=(1-\rho)^{1.9},\
q_y(\rho)=(1-\rho^{1.2})^{1.9},\ q(\rho)=(1-\rho)^{2.5}$.

\subsection{Toward dilute granular flows - granular temperature
revisited}
While the above fittings have been made for the regime of a slow dense
flow with $\nu\to\nu_c$, it is tempting to generalize the theoretical
model so it smoothly crosses over to the standard kinetic continuum
theory (\ref{mass})-(\ref{eps}) for $\rho\to 0$.  This goal can be
achieved by including the equation of state (\ref{state}) and the
equation for the granular temperature (\ref{energy}) back into the
theory. The important difference with respect to the standard kinetic
theory is that the pressure which we calculate with Eq.(\ref{state}) is
not the total pressure, but the partial pressure associated with the
fluid part of the stress tensor. Of course, as $\rho\to 0 $, the static
part of the stress tensor disappears, and the partial pressure becomes
the total pressure. 

\begin{figure}[ptb]
\includegraphics[angle=0,width=2.5in]{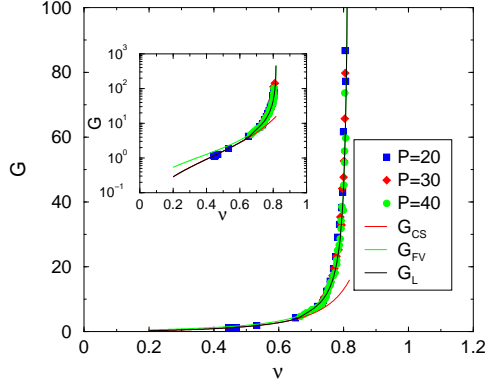}
\vskip 0.5cm
\caption{Particle-particle correlation function $G(\nu)$ calculated via
equation of state (\protect\ref{state}) using the values of fluid 
pressure, temperature, and density in a thin granular Couette flow at
three different external pressures. Inset: the same data in a semi-log
scale
}
\label{kin}
\end{figure}

We can test the relevance of this combined approach by calculating the
spatial correlation function $G(\nu)=(1+e)^{-1}[\pi d^2 p_f/4\nu
T-1]$ with the values of fluid pressure $p_f$, temperature $T$ and
density $\nu$ calculated in our testbed Couette flow at different
external pressures $P$ and comparing it with the theoretical functions
$G_{CS}(\nu), G_{CS}(\nu), G_{L}(\nu)$ (\ref{gCS})-(\ref{gL}). Figure
\ref{kin} shows that the Carnahan-Sterling formula works very well in
the dilute range $\nu<0.67$ as expected. In the high density regime
$G(\nu)$ approaches the free volume result (\ref{gFV}), and the overall
dependence is in agreement with the global interpolation
$G_L(\nu)$ by Luding \cite{Luding01}. 

We can carry this analysis one step further and test the kinetic theory
prediction for the shear viscosity. If we scale $\sigma^f_{yx}$ by 
the shear viscosity calculated using the kinetic formula 
(\ref{shear}) with globally fitted correlation function $G_L(\nu)$ and
actual temperature and density values from corresponding runs, all three
lines in Figure \ref{visc}a  collapse onto the same straight line dependence
$\mu_f^{-1}\sigma^f_{yx}=\dot\gamma$ (see Figure \ref{visc}a, inset).

\begin{figure}[ptb]
\includegraphics[angle=0,width=2.5in]{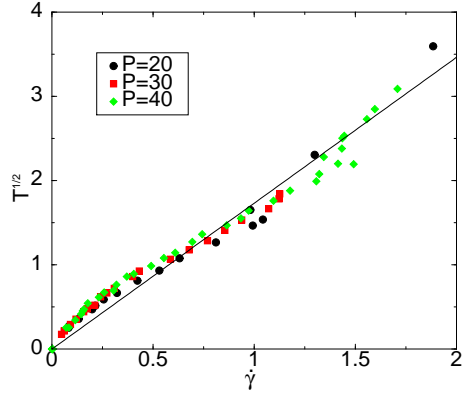}
\vskip 0.5cm
\caption{Granular temperature as a function of strain rate 
in a thin granular Couette flow at three different external pressures. 
}
\label{t_gamma}
\end{figure}

One more test of the kinetic theory predictions can be performed by analyzing the
granular temperature as a function of the shear strain rate. According
to Ref. \cite{savage98}, $T^{1/2}=A\dot\gamma$ for a
plane parallel shear flow where $A$ is a material constant
weakly dependent on volume fraction. Our numerical data for three
different external pressures shown in Figure \ref{t_gamma}, are consistent
with this scaling law although small deviations can be observed at small
$\dot\gamma$. 

We have not done a similar comparison for the bulk viscosity,
thermal conductivity, and the energy loss, however the presented
data strongly suggest that the constitutive
relations for the {\em fluid} part of the stress tensor are well
described by the standard granular hydrodynamics. 

\subsection{Order parameter description of partially fluidized granular
flows - Take 2
\label{take2}} 

Let us summarize the equations of the continuum theory as specified
on the basis of the 2D molecular dynamics simulation of the thin Couette
flow.

The mass, momentum, and energy conservation conditions are expressed by Eqs. 
(\ref{mass})-(\ref{energy}). The order parameter equation now has the form
\begin{equation}
\frac{D\rho}{Dt}=D\nabla^2\rho-(\rho-1)\left(\rho^2-2\rho_*\rho+
\rho_*^2\exp[-A(\delta^2-\delta_*^2)]\right)
\label{GL1}
\end{equation}
with $\rho_*=0.6, \delta_*=0.26, A=25$. As mentioned before, for the
lack of simulation data we assume linear diffusion coupling of the
order parameter with a constant non-dimensional diffusion coefficient $D$.
The constitutive relation now reads
\begin{equation}
\sigma_{\alpha \beta}=[p_f/q_\alpha(\rho)+(\mu_f-\lambda_f)\mbox{Tr}\dot\bgamma/q(\rho)]
\delta_{\alpha \beta}-\mu_f\dot\gamma_{\alpha \beta}/q(\rho)
\label{constit1a}
\end{equation}
with $q_x(\rho)=(1-\rho)^{1.9},\ q_y(\rho)=(1-\rho^{1.2})^{1.9},\ 
q(\rho)=(1-\rho)^{2.5}$.

The equation of state and expressions for viscosity, thermoconductivity
and the energy dissipation have the same functional form as
(\ref{state})-(\ref{eps}), but they are now written for the 
``fluid'' parameters $p_f, \mu_f, \lambda_f$, and $\varepsilon$.

\section{Surface-driven shear granular flow 
under gravity
\label{sec:deep}} 

\begin{figure}[ptb]
\includegraphics[angle=0,width=3.in]{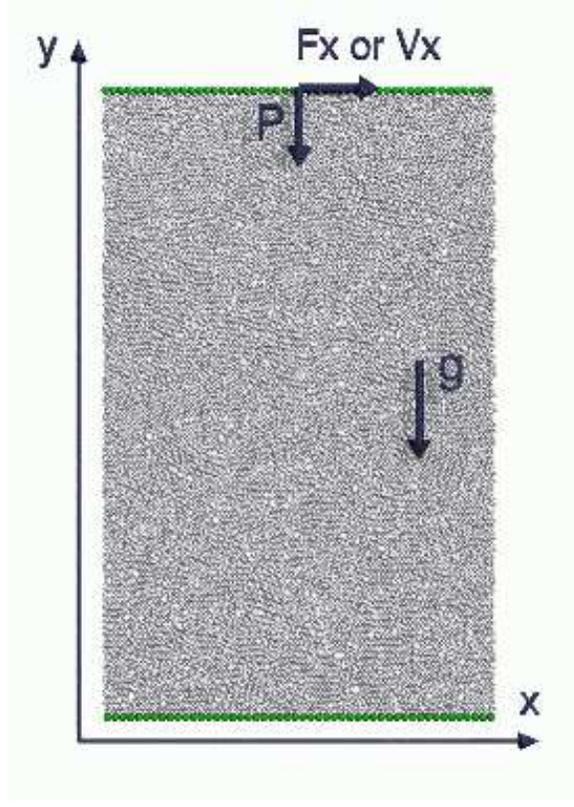}
\vskip 0.5cm
\caption{The geometry of the MD simulation of a surface-driven shear flow} 
\label{setup_thick}
\end{figure}

In this Section we apply the theoretical description which was
formulated in the previous Section on the basis of numerical simulations of a
thin Couette flow with no gravity to {\em another} model problem. We
consider 
shear granular flow in a thick granular layer under gravity
driven by the upper plate which is pulled in a horizontal direction,
see
Figure \ref{setup_thick}. A similar system has been studied
experimentally by Nasuno et al. \cite{gollub98}, as well as by Tsai et
al. \cite{Tsai02}. 

\begin{table}
\begin{center}
\begin{tabular}{lccccccc}
Run ID & $N_p$ & $Lx$ & $Ly$ & $P_e$ & $V_{t}$ & $F_{t}/(P_e L_x)$ & $-\sigma_{yx}$\\ \hline
P10V5   & $5\cdot10^{3}$ & 50 & 100  & 10  & 5   & -  & 5.0 \\
P10V50  & $5\cdot10^{3}$ & 50 & 100  & 10  & 50  & -  & 6.0 \\
P50V5   & $5\cdot10^{3}$ & 50 & 100  & 50  & 5   & -  & 17.0 \\
P50V50  & $5\cdot10^{3}$ & 50 & 100  & 50  & 50  & -  & 25.0 \\
P50V5L  & $1\cdot10^{4}$ & 50 & 200  & 50  & 5   & -  & 18.5 \\
P50V50L & $1\cdot10^{4}$ & 50 & 200  & 50  & 50  & -  & 26.5 \\
P20F10  & $5\cdot10^{3}$ & 50 & 100  & 20  & -   & 10 & 20.0 \\
P20F20XL  & $1.92\cdot10^{4}$ & 96 & 200  & 20  & -   & 20 & 20.0 \\
\end{tabular}
\caption{Parameter values for the simulations of
deep Couette flows
for different geometries and boundary conditions. The first six runs
were performed at  constant velocity of the top plate and the last two
runs were
for a constant horizontal force applied to the plate; for all runs
$\mu_0=0.3,\ e=0.82,\  \Delta_r=0.2,\ k_n=2\cdot10^{5},\ k_s/k_n=1/3,\  
\gamma_n=16.7$.}
\label{table1}
\end{center}
\end{table}

\begin{figure}[ptb]
\includegraphics[angle=0,width=2.5in]{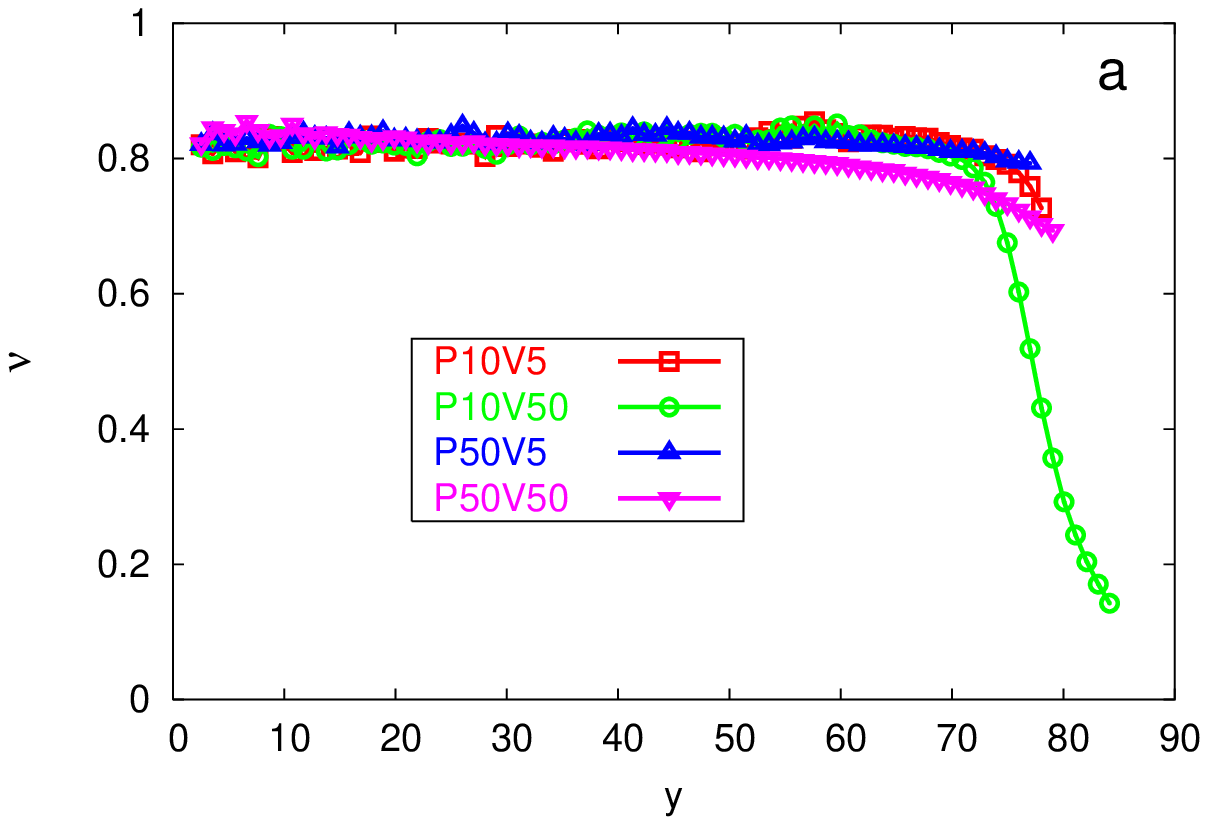}
\includegraphics[angle=0,width=2.5in]{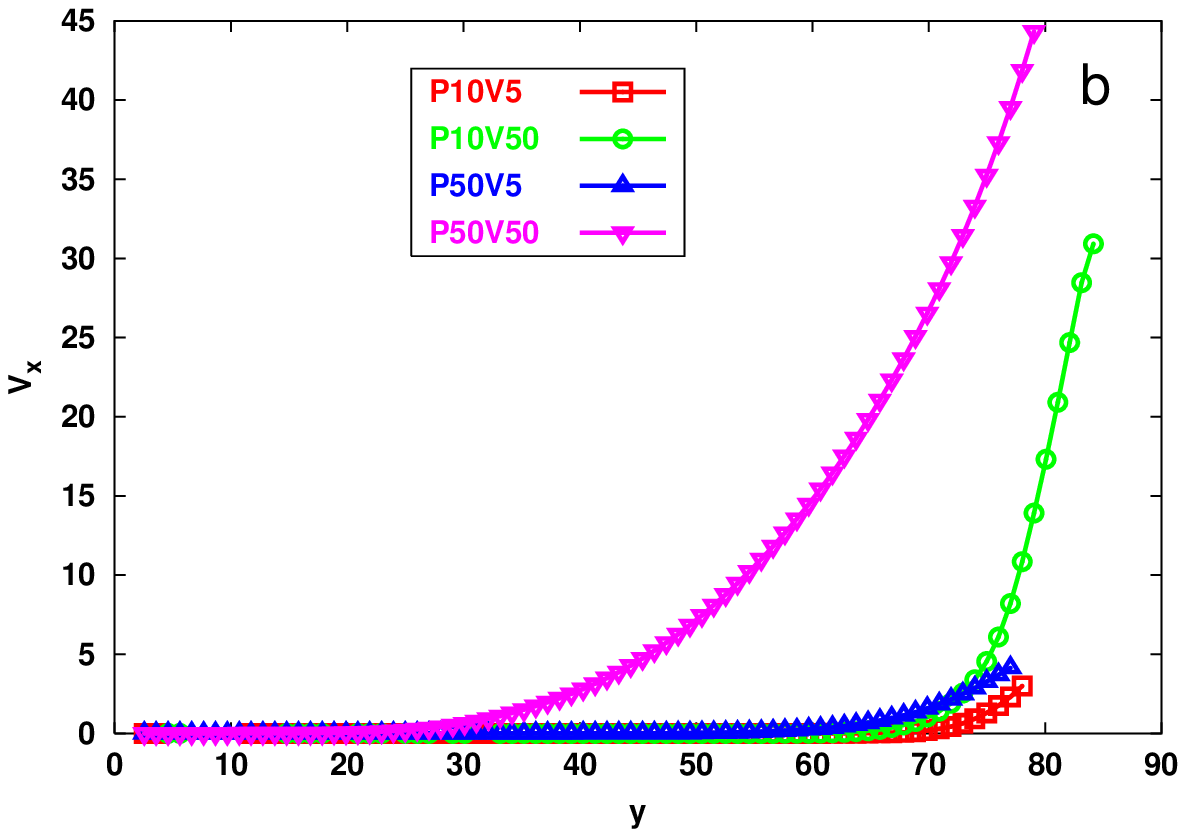}
\includegraphics[angle=0,width=2.5in]{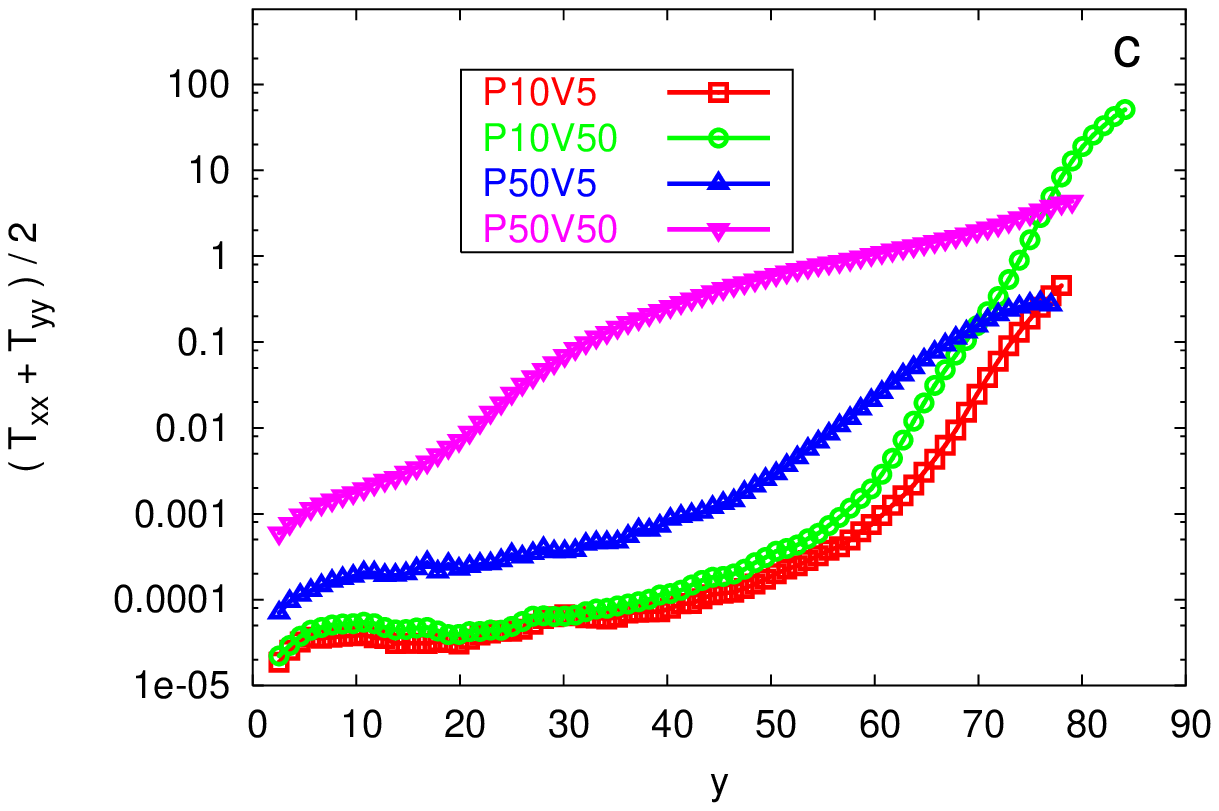}
\includegraphics[angle=0,width=2.5in]{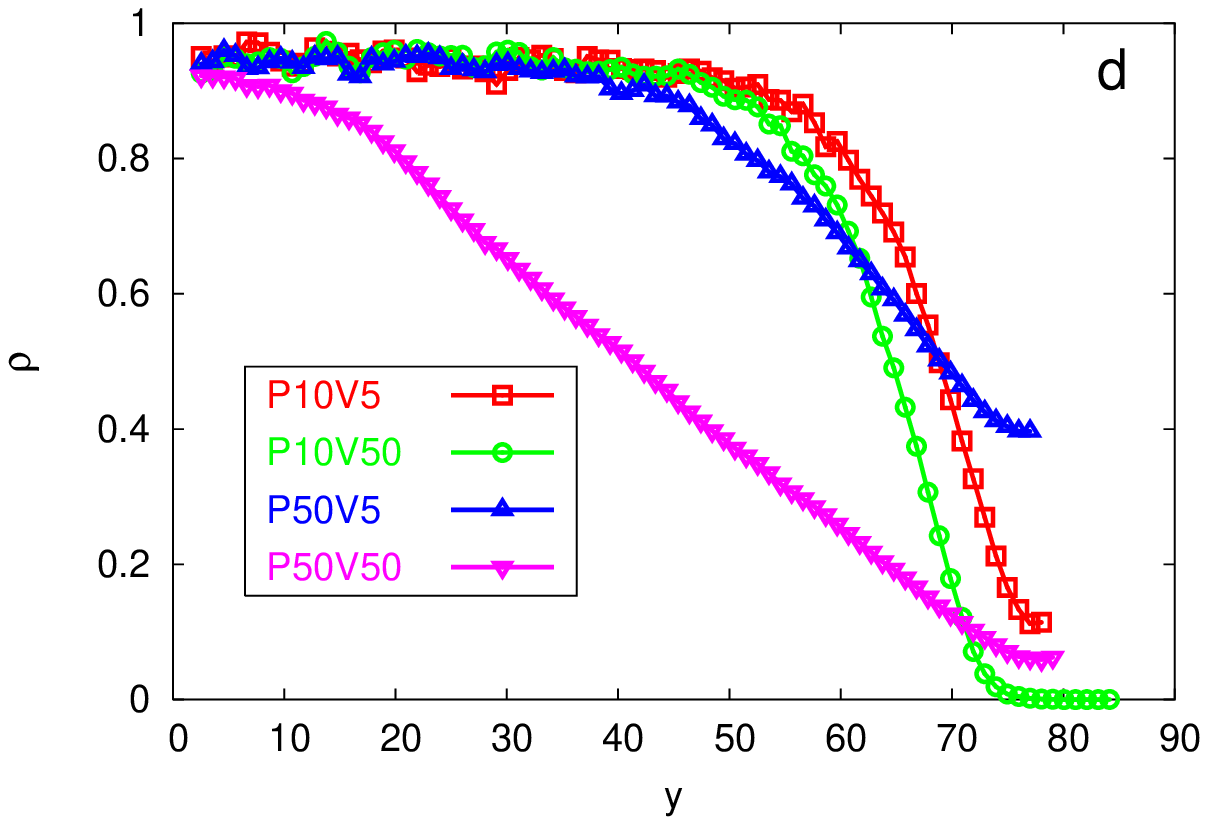}
\vskip 0.5cm
\caption{Density (a), horizontal velocity (b), temperature (c), 
and the order parameter (d) 
profiles in a deep granular layer driven by upper moving plate for four
different runs from Table \protect\ref{table1}.
}
\label{profs}
\end{figure}

We simulated up to 20,000 particles in a rectangular box under a heavy
plate which was moved either with a constant speed $V$ or a constant
force $F$.  Periodic boundary conditions were assumed in a horizontal
direction. After a transient, a quasi-stationary fluidization and shear
flow established in the near-surface layer, while near the bottom grains
remained in a nearly static jammed regime. Here, we show the results of
several different runs with small/large pressure and small/large shear.
The details of these runs are given in Table \ref{table1}.  The vertical
profiles of the density, flow velocity,  and the order parameter are
shown in  Figure  \ref{profs}. The density of grains remain nearly
constant very close to the maximum random packing density value, except
for a narrow near-surface boundary layer. The most significant density
variations were observed for the case $P10V50$ corresponding to a light
fast moving upper plate. The horizontal velocity decays roughly
exponentially off the plate in agreement with experimental evidence
\cite{gollub98,Tsai02}. The vertical profiles of the order parameter
demonstrate a well-defined transition from fluid state near the upper
plate to solid state below. The width of the ``interface'' grows with
the applied pressure $P$, which indicates the stress dependence of the
diffusion coefficient for the order parameter. Surprisingly, we found
that the order parameter does not approach 1 at large depths, but
instead seems to saturate at some value slightly below 1. We believe
that this behavior is inherent to our 2D geometry with periodic boundary
conditions on side walls. The moving upper plate oscillates vertically
and produces vibrations in the bulk of granular layer.  These slowly
decaying with depth vibrations break weak contacts between particles
which are not strongly pressed against each other (e.g. lying under
arches). We believe that in full 3D simulations with more realistic
boundary conditions this effect may be less pronounced. In principle, it
may be included in the theoretical description by proper averaging of
fluctuations of the strain tensor in the spirit of Savage
\cite{savage98}.

\begin{figure}[ptb]
\includegraphics[angle=0,width=2.5in]{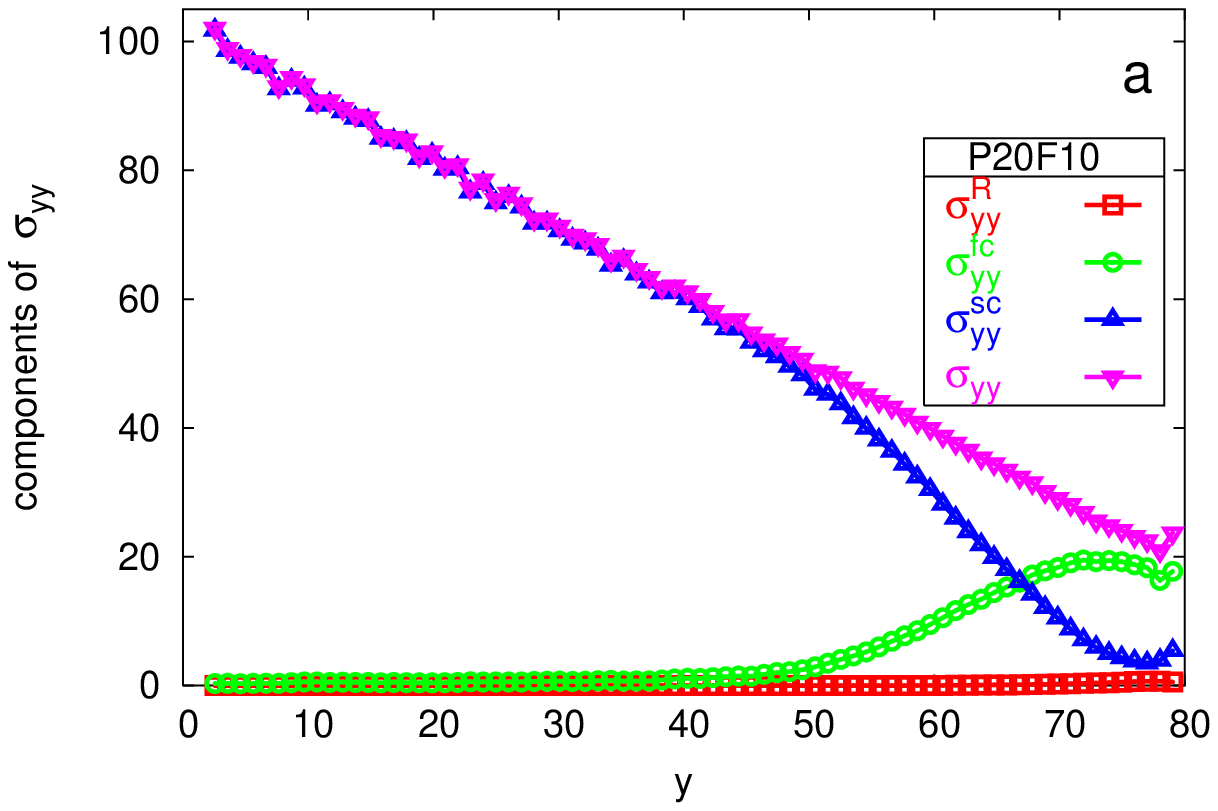}\\
\includegraphics[angle=0,width=2.5in]{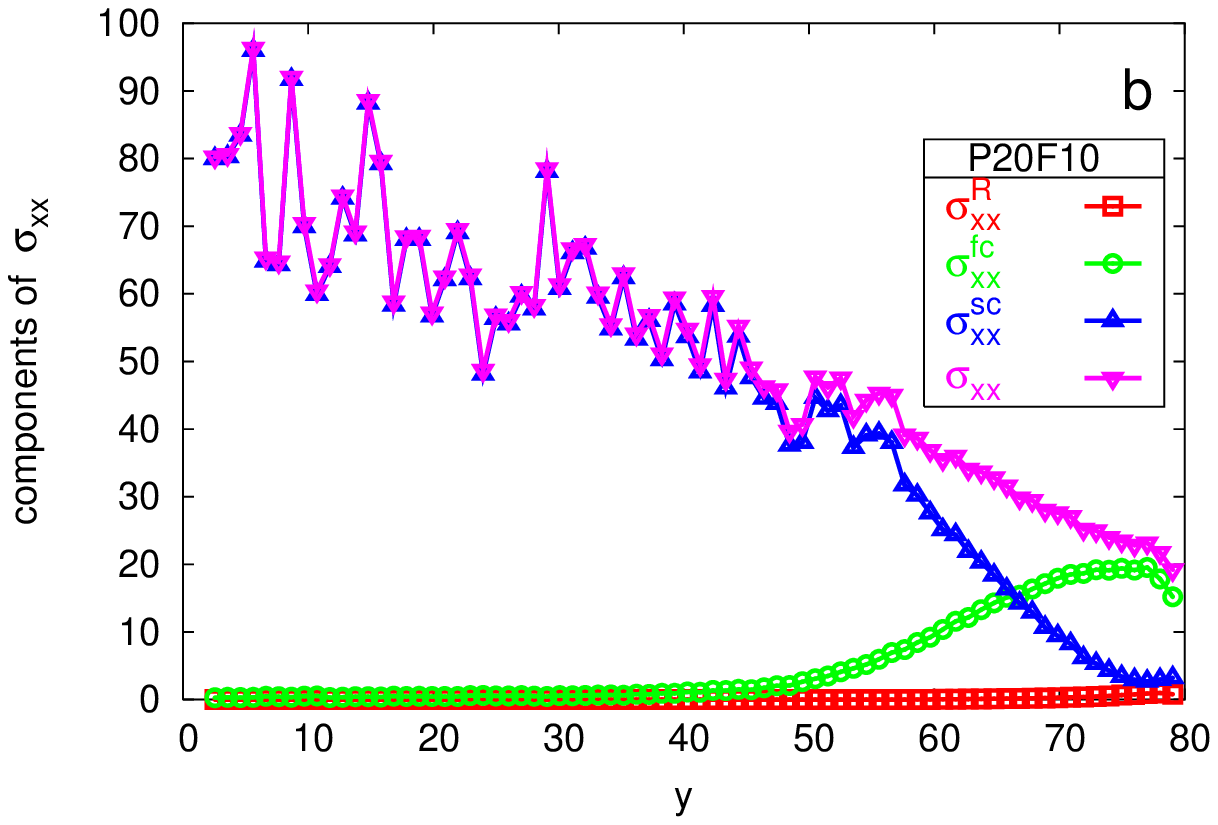}\\
\includegraphics[angle=0,width=2.5in]{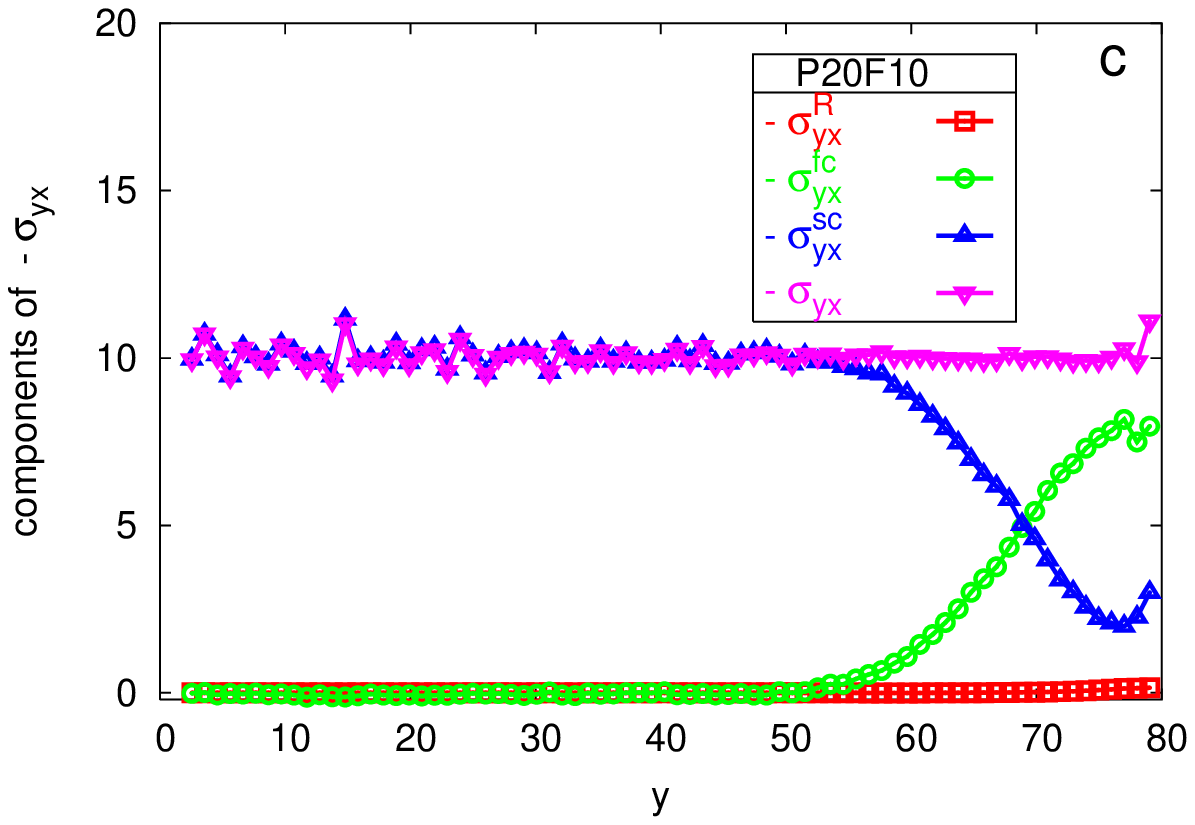}
\vskip 0.5cm
\caption{Stationary profiles of the vertical (a), horizontal (b), and shear (c) stress 
components for run P20F10.}
\label{gravshear_syxprof}
\end{figure}

\begin{figure}[ptb]
\includegraphics[angle=0,width=3.5in]{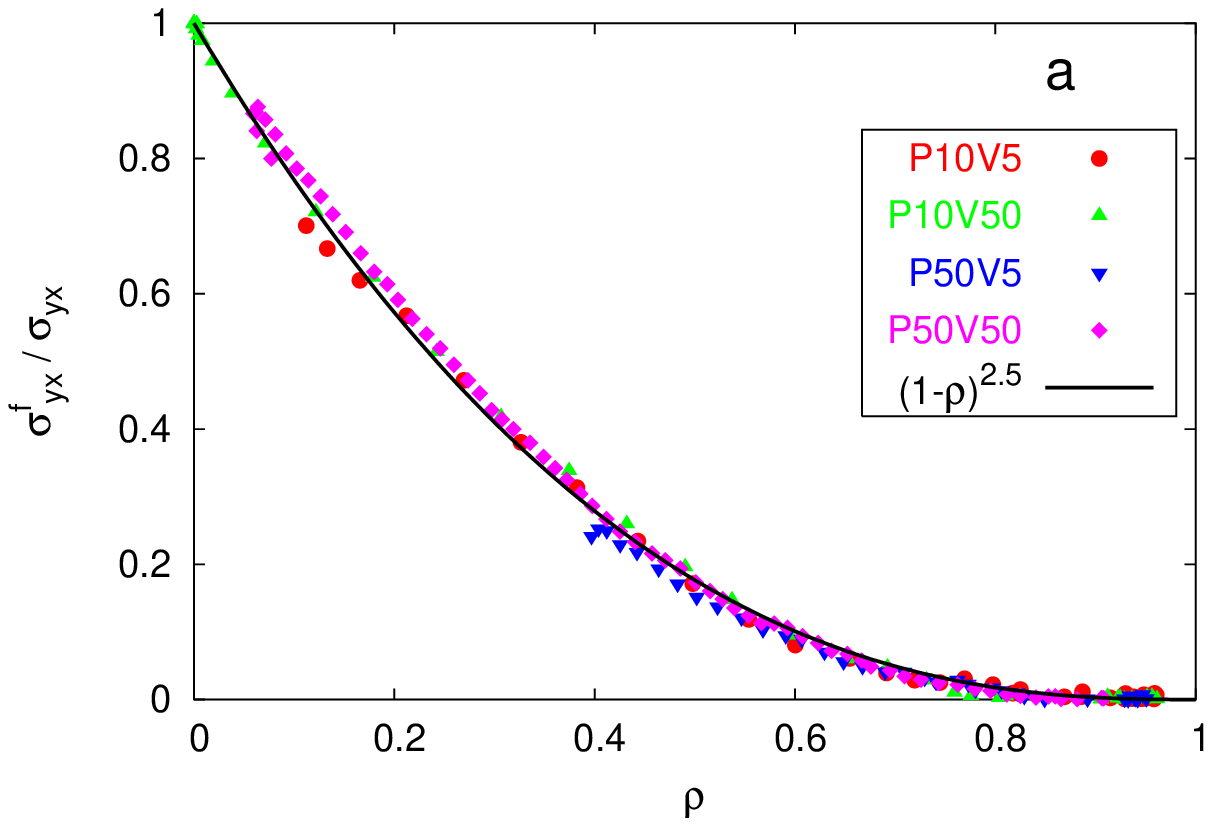}
\includegraphics[angle=0,width=3.5in]{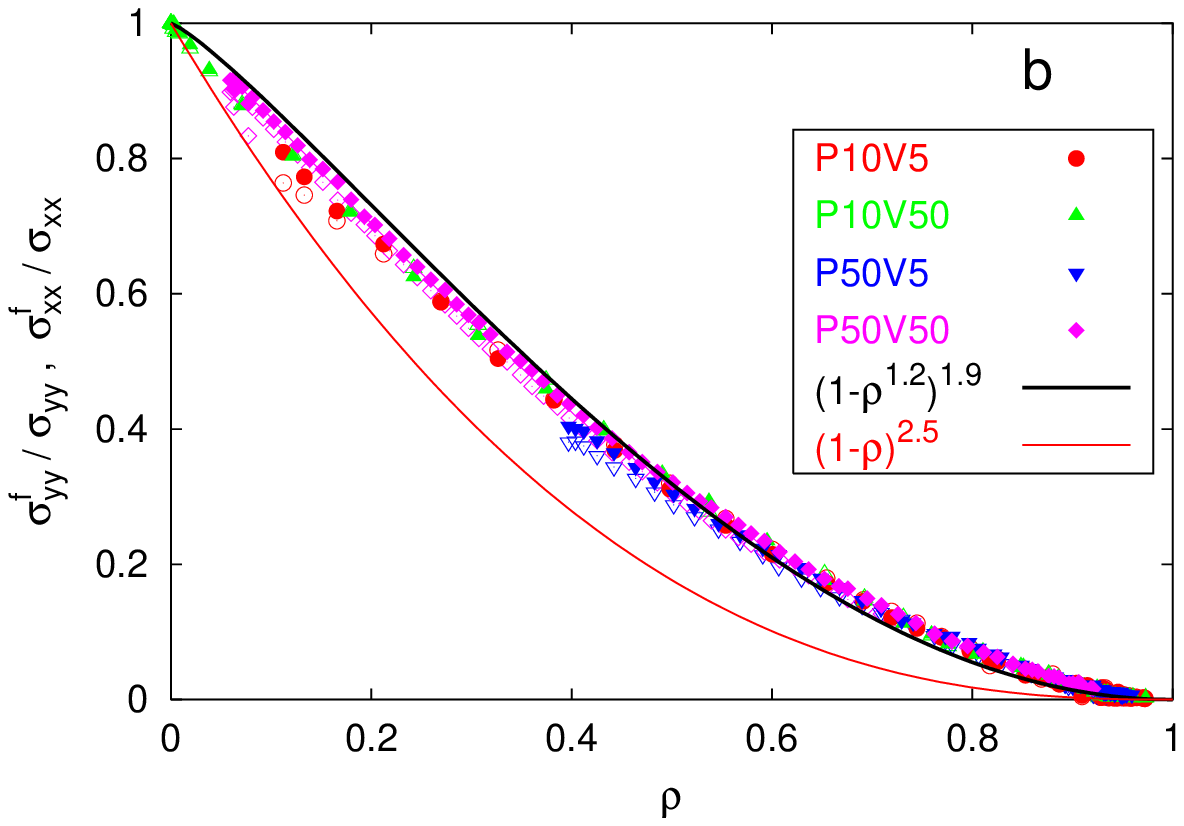}
\vskip 0.5cm
\caption{Ratios of fluid and full components of the stress tensor as a
function of the order parameter for four different runs
at different speeds and pressures: a - shear stress
component $\sigma_{yx}$, b - normal stress components $\sigma_{xx,yy}$. 
Closed symbols correspond to $\sigma_{xx}$, open symbols correspond to 
$\sigma_{yy}$. Solid lines show the fits $q(\rho)=(1-\rho)^{2.5}$ in (a), 
and $q_1(\rho)=(1-\rho^{1.2})^{1.9}$ in (b).
}
\label{StressRatio}
\end{figure}

By averaging velocity fluctuations and forces acting on
individual particles, we calculated vertical profiles of the fluid
and solid parts of the shear and normal stress components
$\sigma_{\alpha \beta}$
(see Section \ref{sec:mds}). These
profiles for Run P10V50 are shown in Figure \ref{gravshear_syxprof}. Strong
fluctuations of the horizontal component of the stress tensor
$\sigma_{xx}$ are mostly related to the static component of the tensor.  According
to Eqs.  (\ref{sigmaf1}),(\ref{sigmas}), fluid and solid parts of the shear 
stress components should be
related to the shear component of the full stress tensor (which in this
geometry is roughly independent of $y$) via the function $q(\rho)$.
Figure \ref{StressRatio},a depicts this function as a parametric plot of
$\sigma^f_{yx}(y)/\sigma_{yx}$ vs.  $\rho(y)$ made using vertical profiles of
stresses and the order parameter.
As seen from this Figure, the same fit $q(\rho)=(1-\rho)^{2.5}$
approximates the data quite well. However, unlike the zero-gravity 
case of the thin Couette flow, the normal stress components seem to be
isotropic $\sigma_{xx}=\sigma_{yy}$ and they both are well described by
$q_y(\rho)$  (see Figure \ref{StressRatio},b).

\begin{figure}[ptb]
\includegraphics[angle=0,width=3.5in]{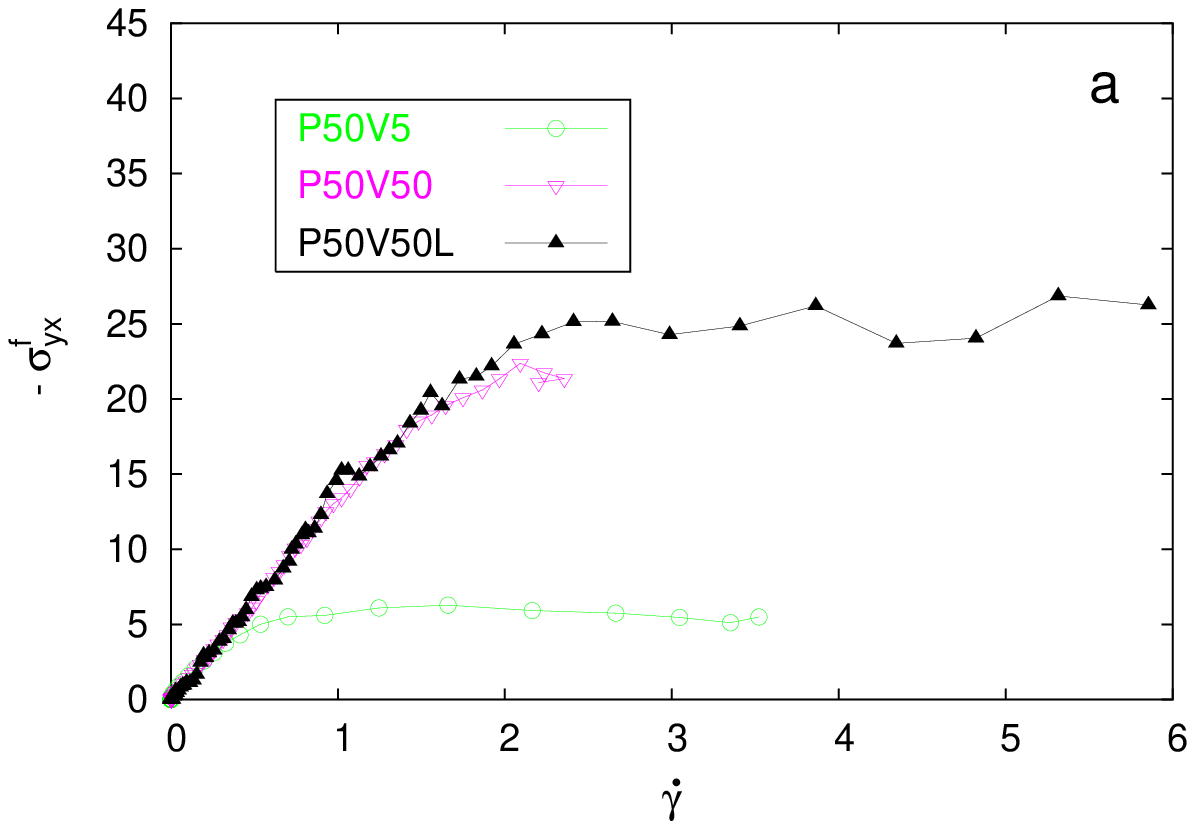}\\
\includegraphics[angle=0,width=3.5in]{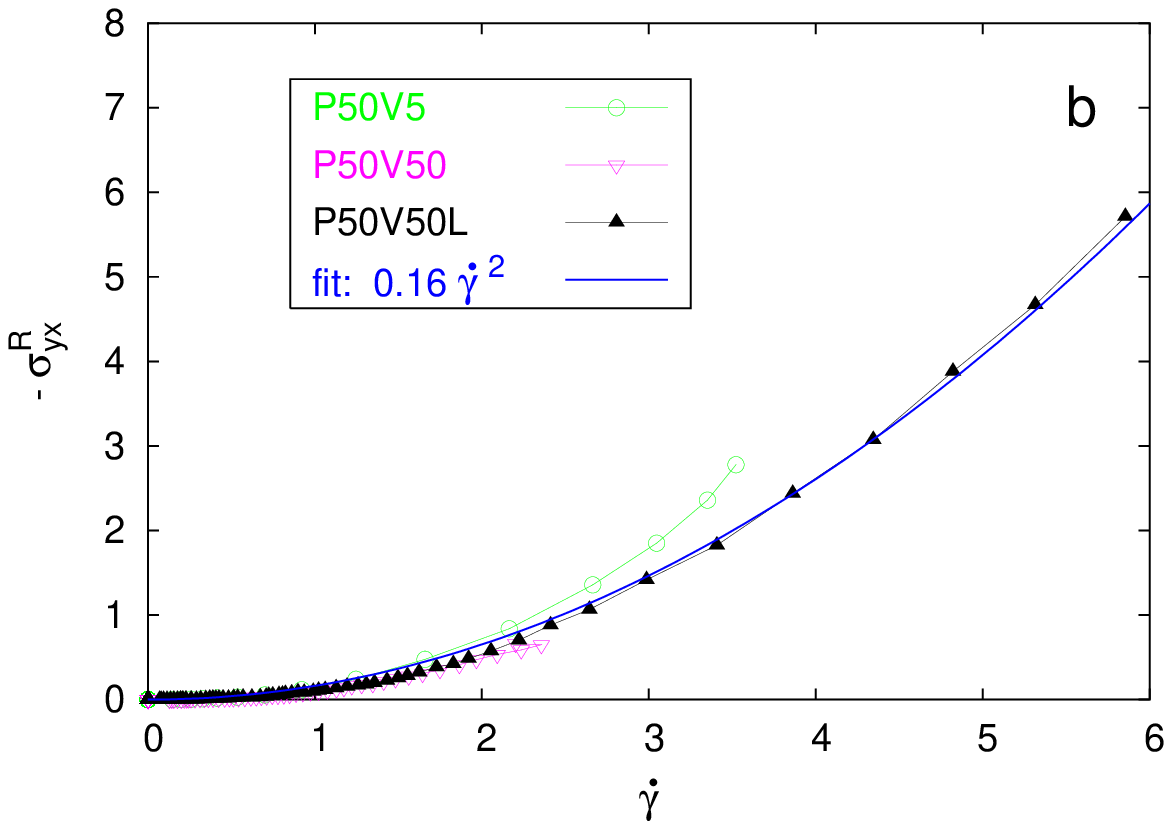}
\vskip 0.5cm
\caption{``Fluid" shear stress vs. local shear strain rate
for several runs: (a) total fluid shear stress, (b) Reynolds
part of the shear stress. }
\label{FSyx-Strain}
\end{figure}

\begin{figure}[ptb]
\includegraphics[angle=0,width=3.5in]{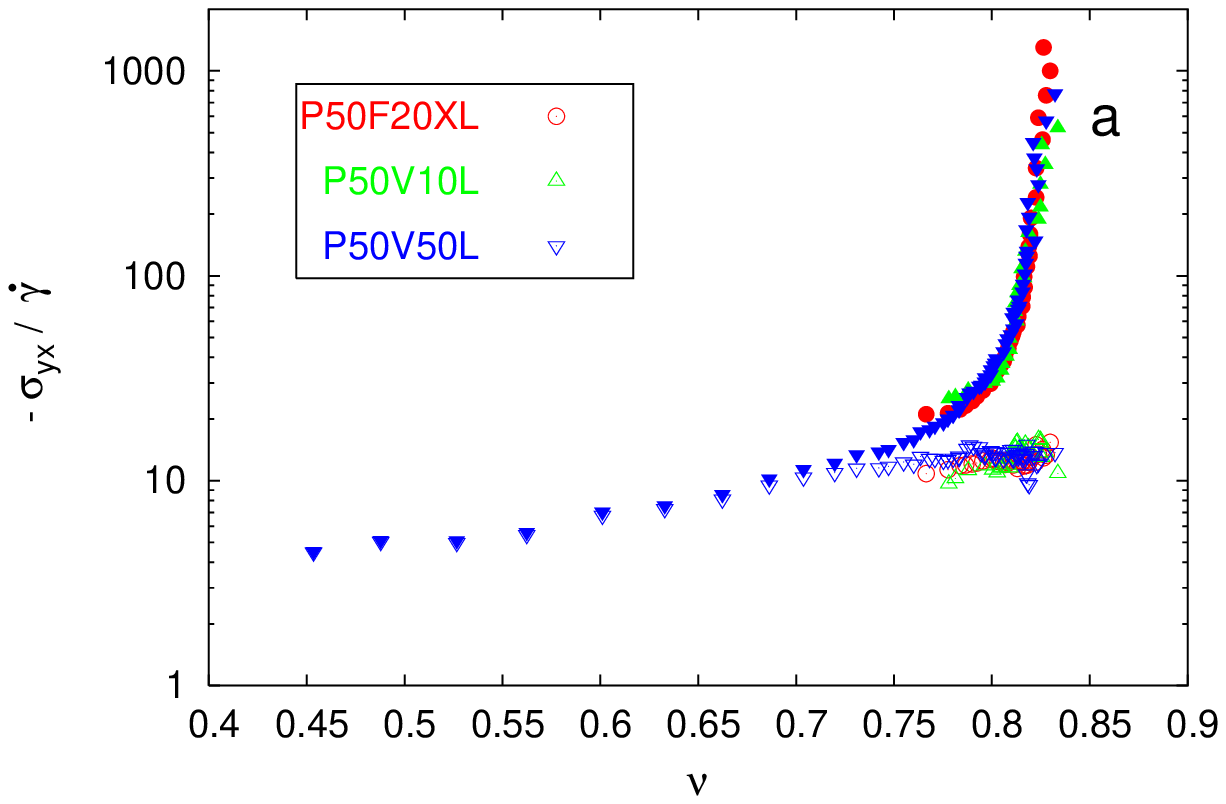}
\includegraphics[angle=0,width=3.5in]{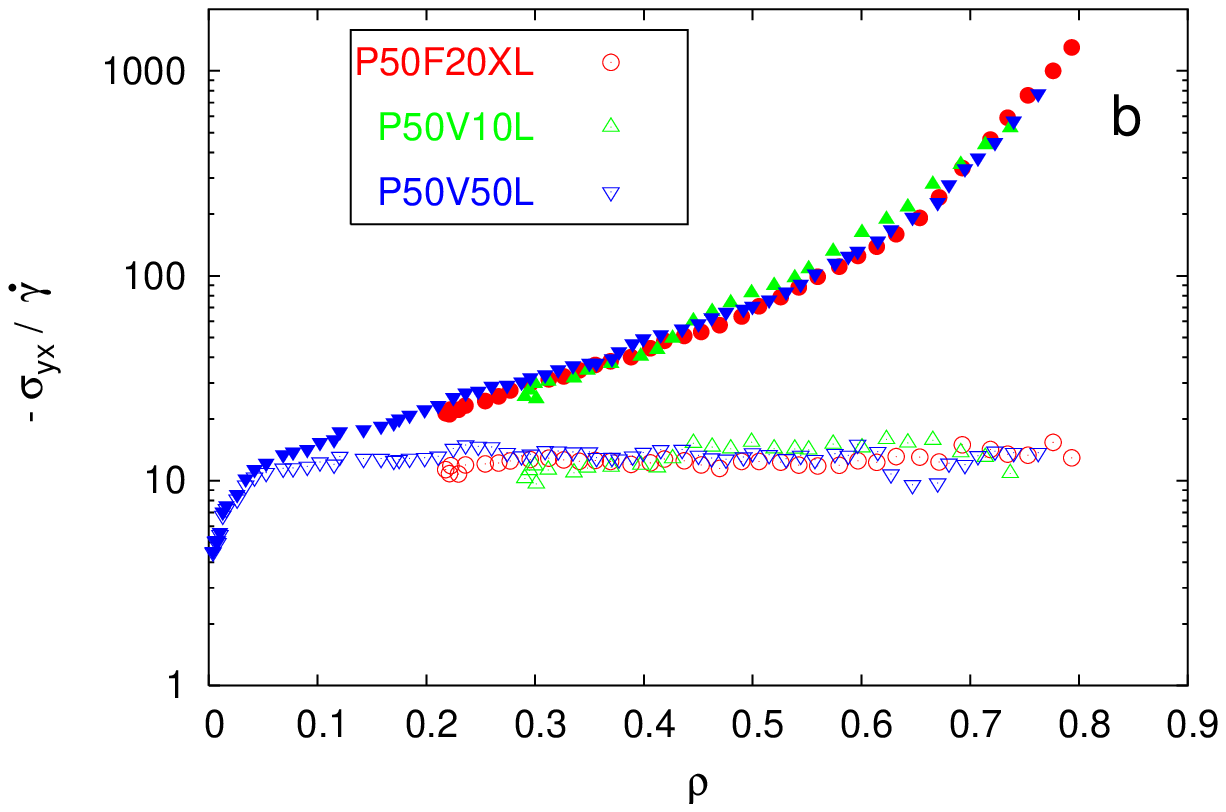}
\vskip 0.5cm
\caption{Full viscosity $-\sigma_{yx}/\dot \gamma$ (solid symbols) and ``fluid'' 
viscosity $-\sigma^f_{yx}/\dot \gamma$  (open symbols) 
coefficients as functions of density (a) and the order parameter (b) for two runs.}
\label{KinTest}
\end{figure}

The dependence of the fluid part of the shear stress component on the
strain rate (Figure \ref{FSyx-Strain}) shows the same behavior as for the 
thin Couette system: at small $\dot\gamma$ the viscosity is nearly
constant $\mu_f\approx 12$, and at larger $\dot\gamma$ shear thinning is
observed. Interestingly, the dependence of the local Reynolds 
shear stress on the local strain rate is well
described by the Bagnold scaling $\sigma^R_{yx}\propto \dot\gamma_{yx}^2$.
Eventually, at large $\dot\gamma$, this scaling should dominate the
full stress-strain rate relationship.
Figure \ref{KinTest} compares the behavior of the viscosity coefficient
$\mu=\sigma_{yx}/\dot\gamma$ and $\mu_f=\sigma^f_{yx}/\dot\gamma$
calculated along the vertical profiles of $\sigma$ and $\dot\gamma$, as a
function of density $\nu$ and order parameter $\rho$. While the former
diverges as $\nu\to\nu_c$ and $\rho\to 1$, the latter approaches the
constant value $\mu_f\approx 12$, in agreement with results of Section
\ref{sec:testbed}.

\begin{figure}[ptb]
\includegraphics[angle=0,width=3.5in]{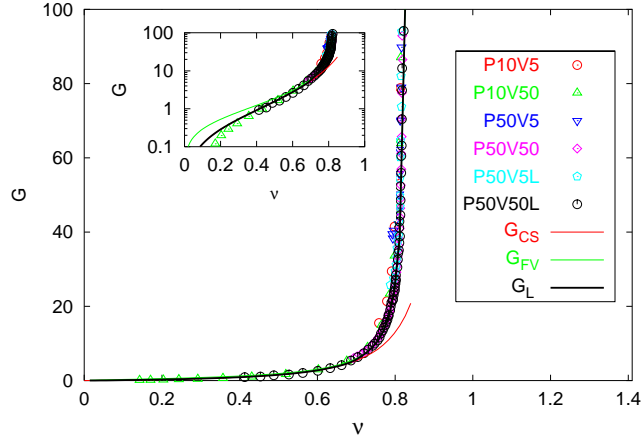}
\vskip 0.5cm
\caption{Particle-particle correlation function as a function 
of density for several runs of the thick Couette flow simulations}
\label{Gcs_KinTest}
\end{figure}

As in the previous Section, we can extract the particle-particle
correlation function by calculating $G(\nu)=(1+e)^{-1}[\pi d^2
p_f/4\nu T-1]$ using the vertical profiles of $\sigma^f_{yx}, T$, and
$\nu$.  Again, we obtain a good agreement with theoretical predictions
based on the kinetic theory for fluid component of the stress tensor
(Figure \ref{Gcs_KinTest}).

Finally, we can compare the stationary vertical profiles of the order
parameter and the horizontal velocity with theoretical predictions. In
most of our numerical simulations we specified the velocity of the upper
plate rather than the applied force. That allowed us to study the
regimes of slow dense flows which would be unstable had we applied a
constant shear force.  The shear stress tensor component $\sigma_{yx}$
in the stationary regime was indeed constant across the layer (see for
example Figure\ref{gravshear_syxprof},c).  However, due to slippage near
the moving plate the relation between the plate speed and the shear
stress is complicated. We do not address the issue of boundary
conditions here as it is a subject of a separate study (see for example
\cite{hui84}). Here we simply use the values which are obtained in
numerical simulations (the last column in Table \ref{table1}), as
parameters in our theoretical model.

In the stationary regime, the relevant stress tensor components are 
specified as follows:
\begin{equation}
\sigma_{yy}=P+(H-y),
\end{equation}
\begin{equation}
\sigma_{yx}=\sigma_{xy}=const
\end{equation}
where $P$ is the external pressure applied to the upper wall.  

We also need to specify the boundary conditions for the order parameter
at the top and bottom plates. 
This is a serious issue in its own right which we will
address elsewhere. Here we simply impose no-flux boundary conditions
both at the top and the bottom plate for the order parameter,
$\partial_y\rho(0)=\partial_y\rho(H)=0$.  

We limit ourselves with the case of slow dense flow regime, when the
granular temperature plays a minor role, and the granular flow can be
considered incompressible. This allows us to use the reduced set of
equations (\ref{momentum}),(\ref{GL1}),(\ref{constit1a}) with the fixed viscosity
$\mu_f=12$.  The stationary shear flow solution of the continuum equations
can be found numerically as follows. Since the
components of the full stress tensor are assumed known, we solve the
time-dependent order parameter equation (\ref{GL1}) using the
pseudo-spectral method until the solution reaches a stationary state.
The resulting solution for the order parameter is then used to obtain
the velocity profile by integrating the constitutive relation
(\ref{constit1a}) from the bottom ($y=0$) up. Since the grains are
strongly compressed near the rough bottom plate due to gravity, we
assume the no-slip boundary condition for the horizontal velocity at
$y=0$. The momentum conservation equation (\ref{momentum}) is satisfied
automatically. Thus obtained profiles of velocity and the order parameter were
compared with our 2D molecular dynamics simulations.

\begin{figure}[ptb]
\includegraphics[angle=0,width=3.in]{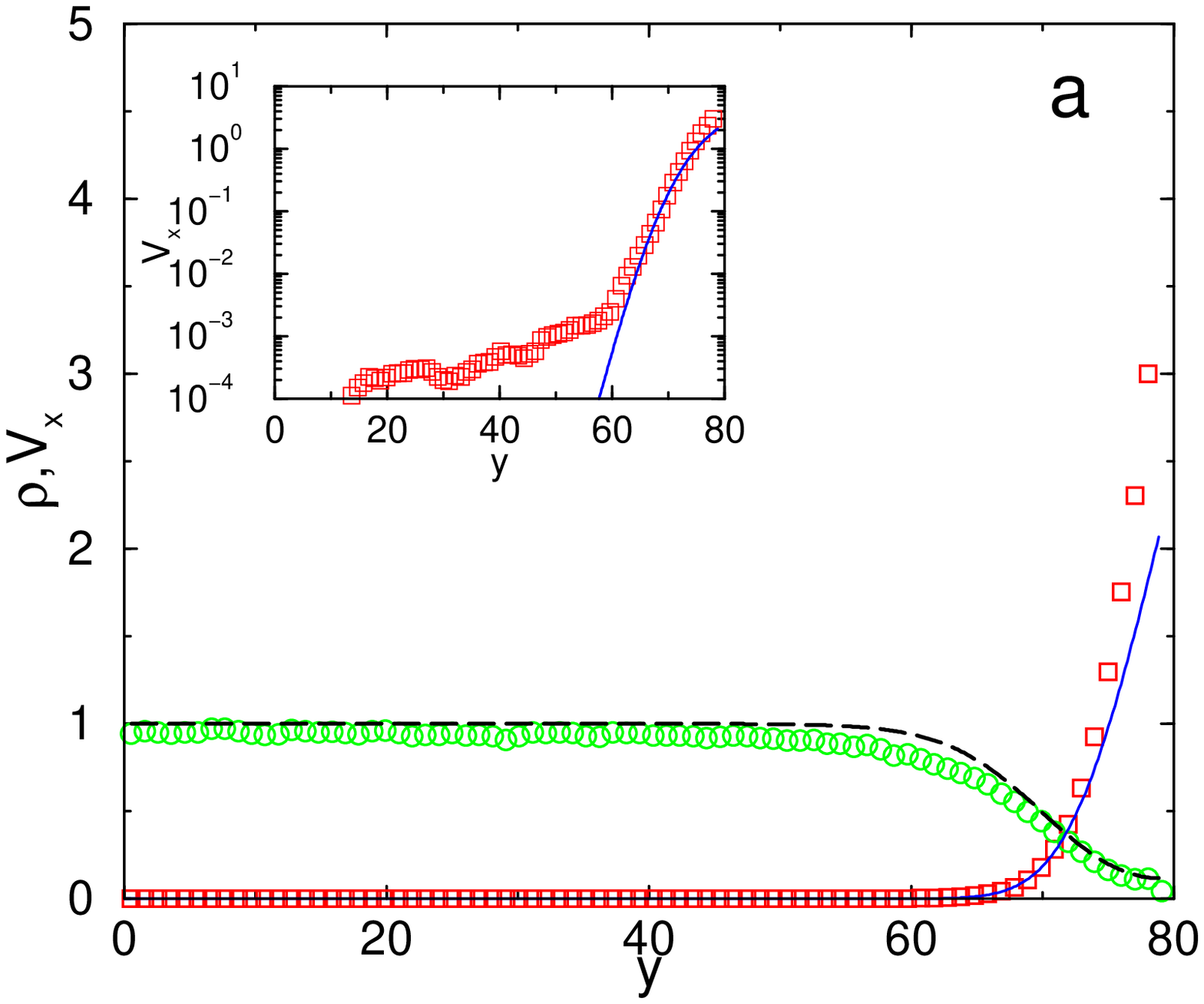}
\includegraphics[angle=0,width=3.in]{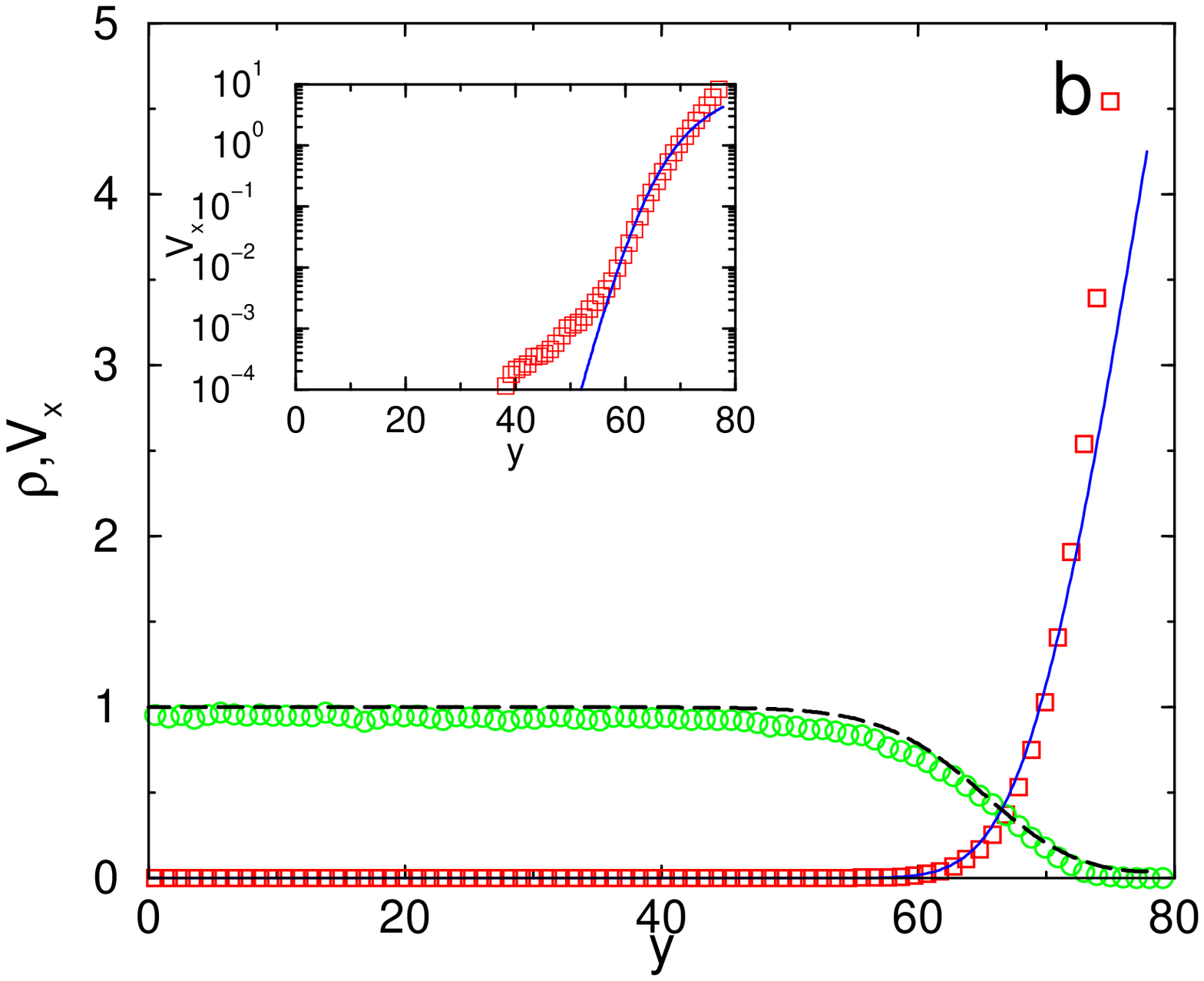}
\includegraphics[angle=0,width=3.in]{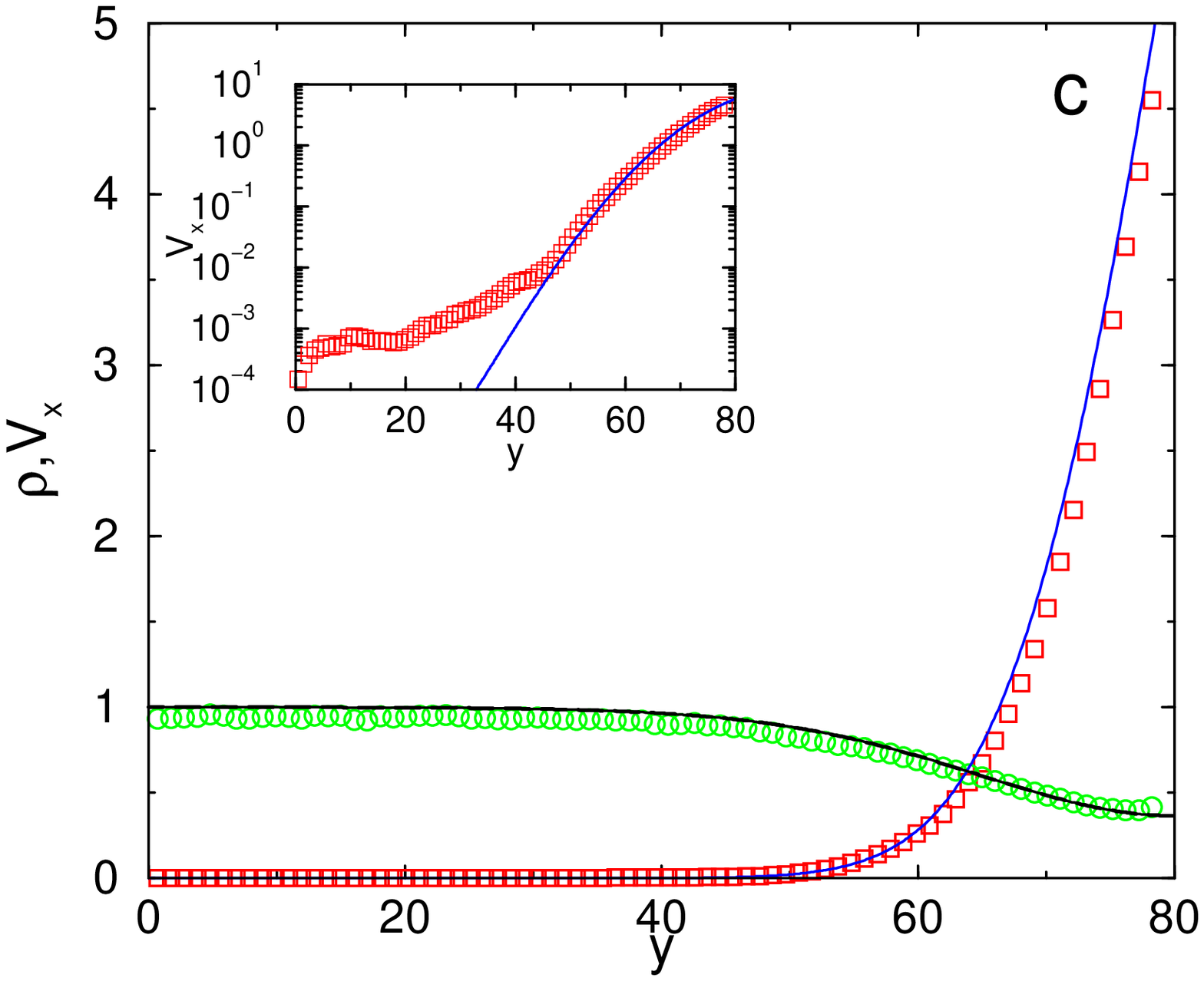}
\includegraphics[angle=0,width=3.in]{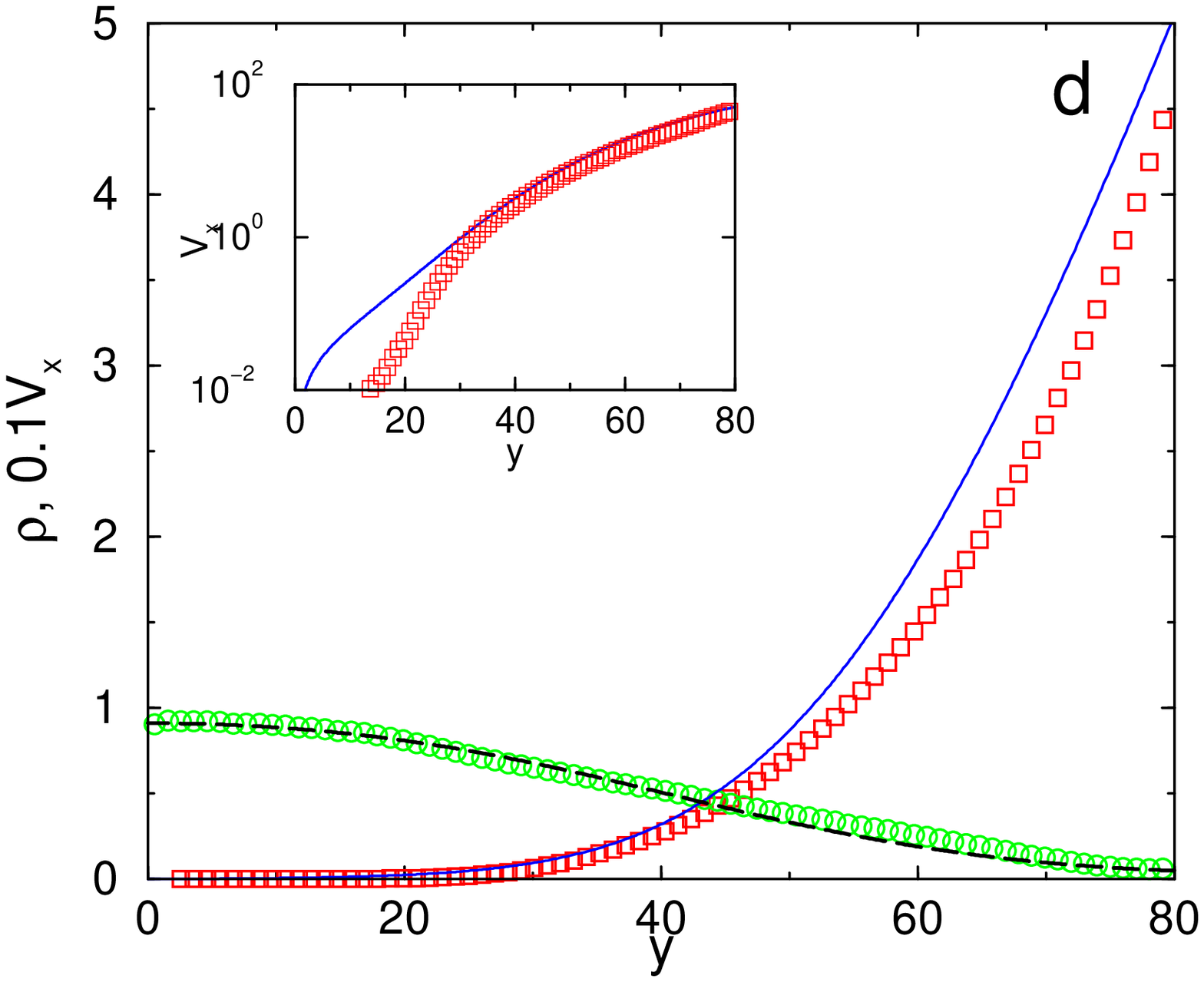}
\vskip 0.5cm
\caption{Profiles of the order parameter and velocity in a thick
granular layer driven at the surface by a heavy moving plate for 
Run P10V5 (a), P10V50 (b), P50V5 (c), P50V50 (d). 
Lines show the theoretical results obtained from the continuum 
model(\protect\ref{GL1}),(\ref{constit1a}), empty symbols indicate numerical
data. Insets show the velocity profiles in the logarithmic scale.}
\label{profiles}
\end{figure}

The results of the comparison between the velocity and order parameter
profiles obtained in simulation and by using the continuum theory are
shown in Figure \ref{profiles} for four runs P10V5 (a), P10V50 (b), 
P50V5 (c), P50V50 (d). The only fitting parameter used was the diffusion
constant $D$ in the order parameter equation, which has not been
determined in our testbed analysis. We used $D=1$ for runs P10V5 and
P10V50, $D=5$ for P50V5 and $D=10$ for P50V50. From this we can conclude
that the diffusion coefficient depends on the 
the local stress tensor, however more
elaborate numerical experiments are needed to pinpoint this dependence
more quantitatively.  All other parameters were identical for all four
cases as specified in Section \ref{take2}.  The vertical profiles of the order parameter 
and the horizontal velocities
are reasonably well described by the theory. However, for low pressure
runs P10V5 and P10V50, the horizontal velocity profiles
deviate from the numerical data presumably
because the viscosity coefficient is no longer a constant in a dilute
region near the top plate.  

\section{Conclusions}  
In this paper we performed a series of numerical simulations of 2D
wall-driven granular flows. These simulations were designed with a specific goal,
 to quantify the continuum theory of partially fluidized
granular flows which was introduced  {\em ad hoc} earlier
\cite{AT1,AT2}.  We defined the order parameter as a ratio of the
number of static contacts to the total coordination number averaged over
a small mesoscopic volume. Using simulations of a
thin Couette flow between two rough plates, we determined
the free energy density for the order parameter. Simulations confirmed
that the ratio of the shear to the normal stress in the bulk of the
granular flow can parametrize the stationary states of the order parameter equation.
The same simulations allowed us to determine the detailed
structure of the constitutive relation. We split the total stress
tensor into the fluid and solid components, in which the former is
comprised of the Reynolds stresses and the stresses transmitted through
short-term collisions, while the latter is formed by the force chains
through persistent contacts. The ratio of fluid and solid stress
components is indeed determined by the order parameter through
scaling functions $q(\rho),\ q_{x,y}(\rho)$.
Remarkably, the fluid component of the stress tensor is a
linear function of the strain rate $\dot\gamma$ in the slow dense flow
regime. This justifies the Newtonian scaling of the stress-strain
relationship adopted in the theory. 

Using the calibrated theory, we studied the flow structure of a thick
surface driven Couette granular flow under gravity. We found the theoretical
predictions to be in a good {\em quantitative} agreement with
simulations. 

The evidence presented here suggests an intriguing interpretation for
the order parameter description of dense and slow granular flows.  The granular
material under shear stress appears to be similar to a multi-phase
system with the fluid phase ``immersed'' into the solid phase.  The
fluid phase behaves as a simple Newtonian fluid for small shear rates
when the density is almost constant, but exhibit shear thinning at
larger shear rates when the density begins to drop. We observed that the
Reynolds part of the fluid shear stress obeys the Bagnold scaling
$\sigma^R_{yx}\sim \dot \gamma^2$. We anticipate that for very large shear
rates when the Reynolds stress becomes dominant the overall stress
tensor should exhibit Bagnold scaling locally.

While this theory is primarily intended for dense
and slow granular flows, we have shown that it can be combined with
existing models of rapid granular flows based on the kinetic theory of
granular gases. This requires to drop the assumption of
incompressibility and include the equation for the granular temperature. 
Our simulations showed that the kinetic theory works well for
the {\em fluid} part of the stress tensor in the {\em whole} range of
densities from dilute regime to the critical random close packing
density. 

Many issues still remain open.  The spatially non-uniform dynamics of
the order parameter requires a more detailed study. We found that the
diffusion constant postulated in Eq.(\ref{GL1}) appears to be a function
of the normal shear stress as well as the local strain rate, however, we
do not have sufficient numerical data to provide a quantitative
description of this dependence. It would be of interest to analyze the
propagation of a fluidization front in a granular layer prepared in a
meta-stable static regime. Such simulations could provide an insight
into the mechanisms of the local coupling of the order parameter. 

The molecular dynamics algorithm employed is based on a number of
approximations. These approximations, however well tested and widely
accepted \cite{wolf96,grest91,grest02}, directly affect the results of
our fitting the continuum model. For example, if one replaces the Hookian
model of particle interaction with a Hertzian one, an appreciable
difference in the structure of the order parameter may be observed. More numerical
work is needed to quantify the relationships between the microscopic
parameters of the system (nature of collisions, restitution coefficient,
friction, etc) and the parameters of the continuum model.

Finally, our simulations were limited by 2D systems, and of course the
resulting continuum theory can only be directly applicable to 2D
systems.  While we anticipate that the structure of the model to
remain in 3D systems, the specific form of the fitting
functions should change. This future work will allow us to perform a
comparison of the 3D model not only with numerical simulations,
but also with experimental data.

The authors are indebted to B. Beringer, P. Cvitanovic, J. Gollub, T. Halsey, 
and J. Vi\~nals
for useful discussions, and to J. C. Tsai for sharing his unpublished
experimental data.  This work was supported by the Office of the Basic
Energy Sciences at the US Department of Energy, grants W-31-109-ENG-38,
and DE-FG03-95ER14516. 
Simulation were performed at the National Energy Research
Scientific Computing Center.


\begin{thebibliography}{99}
\bibitem{note1} This assumption implies that stress tensors $\sigma,\
\sigma^f,\ \sigma^s$ are collinear, which as we see in the following
Sections, is a reasonable assumption for our 2D simulations, however
does not have to hold in all cases. In those circumstances, the order
parameter $\rho$ would become a tensor itself.
\end{thebibliography}
\end{document}